\documentclass[twocolumn,aps,superscriptaddress,showpacs,times]{revtex4}

\usepackage{amssymb}
\usepackage{amsmath}
\usepackage{graphicx}
\usepackage[normalem]{ulem}
\usepackage[dvips]{color}
\begin{document}

\title{Application of microscopic transport model in the study of nuclear equation of state from heavy ion collisions at intermediate energies}

\author{Yongjia Wang}
\affiliation{School of Science, Huzhou University, Huzhou 313000, China}

\author{Qingfeng Li\footnote{Corresponding author: liqf@zjhu.edu.cn}}
\affiliation{School of Science, Huzhou University, Huzhou 313000, China}
\affiliation{Institute of Modern Physics, Chinese Academy of Sciences, Lanzhou 730000, China}

\date{\today}

\begin{abstract}
The equation of state (EOS) of nuclear matter, i.e., the thermodynamic relationship between the binding energy per nucleon, temperature, density, as well as the isospin asymmetry, has been a hot topic in nuclear physics and astrophysics for a long time. The knowledge of the nuclear EOS is essential for studying the properties of nuclei, the structure of neutron stars, the dynamics of heavy ion collision (HIC), as well as neutron star mergers. HIC offers a unique way to create nuclear matter with high density and isospin asymmetry in terrestrial laboratory, but the formed dense nuclear matter exists only for a very short period, one cannot measure the nuclear EOS directly in experiments. Practically, transport models which often incorporate phenomenological potentials as an input are utilized to deduce the EOS from the comparison with the observables measured in laboratory. The ultrarelativistic quantum molecular dynamics (UrQMD) model has been widely employed for investigating HIC from the Fermi energy (40 MeV per nucleon) up to the CERN Large Hadron Collider energies (TeV). With further improvement in the nuclear mean-field potential term, the collision term, and the cluster recognition term of the UrQMD model, the newly measured collective flow and nuclear stopping data of light charged particles by the FOPI Collaboration can be reproduced. In this article we highlight our recent results on the studies of the nuclear EOS and the nuclear symmetry energy with the UrQMD model. New opportunities and challenges in the extraction of the nuclear EOS from transport models and HIC experiments are discussed.

\end{abstract}

\pacs{21.65.Ef, 21.65.Mn, 25.70.-z}

\maketitle
\section{Introduction}

Matter with extremely conditions, such as high density, temperature, and isospin asymmetry, is hardly observed on earth, but can be found at various astrophysical objects. Study of the properties of dense matter may provide deep insight into the structure and evolution of astrophysical objects. The nuclear equation of state (EOS)
which characterizes the thermodynamic relationship between the binding energy $E$, temperature $T$, density $\rho$, as well as the isospin asymmetry $\delta=(\rho_{n}-\rho_{p})/(\rho_{n}+\rho_{p})$ in nuclear matter (an uniform and infinite system with neutrons and protons) has attracted considerable attention from both nuclear physics and astrophysics communities since long time ago\cite{BALi08,Tsang:2012se,Baldo:2016jhp,Oertel:2016bki,Li:2018lpy,Roca-Maza:2018ujj,Burrello:2019wyi,Giuliani:2013ppnp,Ma:2018wtw,Ono:2019jxm,Xu:2019hqg,Chen:2013uua,Xu:2009vi,Gao:2019vby}. At zero temperature, both phenomenological and microscopic model calculations have indicated that the binding energy per nucleon in isospin asymmetric nuclear matter can be well approximated by $E(\rho,\delta)=E(\rho,\delta=0) + E_{sym}(\rho)\delta^{2} + \mathcal{O}(\delta^{4})$. The first term $E(\rho,\delta=0)$ is the binding energy per nucleon in the isospin symmetric nuclear matter, $E_{sym}(\rho)$ is the density-dependent nuclear symmetry energy. Odd-order $\delta$ terms are vanished because of the charge symmetry of nuclear forces. Higher-order terms for $\delta$ are usually negligible for most investigations as the typical value of $\delta$ is about 0.2 in nuclei and nuclear collisions. It is worth noting that higher-order terms in $\delta$ may play an important role in the studying of astrophysical processes\cite{Cai:2011zn,Steiner:2006bx,Pu:2017kjx,Liu:2018far}. The parabolic approximation would be expected to be valid only at small $\delta$, however based on many theoretical calculations, it turns out that it is fairly satisfied even at $\delta$=1 with moderate density\cite{BALi08}.

It is of great interest to investigate how $E(\rho,\delta=0)$ and $E_{sym}(\rho)$ vary as density, because the information of both $E(\rho,\delta=0)$ and $E_{sym}(\rho)$ are essential for studying the structures and the properties of nuclei and neutron stars, the dynamics of heavy-ion collision, supernovae explosions, as well as neutron star mergers. It is also one of the fundamental goals of the current and future nuclear facilities (e.g., the CSR and HIAF in China, the FRIB in the United States, the RIBF in Japan, the SPIRAL2 in France, the FAIR in Germany) around the world. Practically, the nuclear incompressibility $K_0=9\rho^2\left(\frac{\partial^2E(\rho,\delta=0)}{\partial\rho^2}\right)|_{\rho=\rho_{0}}$, the symmetry energy coefficient $S_0=E_{\rm sym}(\rho_{0})$, and its slope parameter $L=3\rho\left(\frac{\partial{E_{\rm sym}(\rho)}}{\partial\rho}\right)|_{\rho=\rho_{0}}$ and its curvature parameter $K_{sym}$=$9\rho^2\left(\frac{\partial^2{E_{sym}(\rho)}}{\partial\rho^2}\right)|_{\rho=\rho_{0}}$  which characterize how $E(\rho,\delta=0)$ and $E_{sym}(\rho)$ change as density, have attracted considerable attention. Although great endeavors have been made to constrain these parameters, a precise picture of the EOS has still not emerged, especially at high densities, which remains an open challenge for further research.

In the present work, we review and highlight our recent results on the studies of the nuclear EOS of isospin symmetric matter, the medium effects on the nucleon-nucleon cross section, and the density-dependent nuclear symmetry energy based on the ultrarelativistic quantum molecular dynamics (UrQMD) model. This article is organized as follows. In next section, the UrQMD model and its recent updates, as well as observables in HIC are briefly introduced. In Section \ref{set1}, the influence of the in-medium nucleon-nucleon cross section on observables in HIC at intermediate energies (with beam energy of several hundreds MeV per nucleon) is discussed. Section \ref{set2} gives the result of studying the nuclear EOS of isospin symmetric matter from the rapidity-dependent elliptic flow. Constraints on the density-dependent nuclear symmetry energy with the UrQMD model are reviewed and discussed in Section \ref{set3}. Finally, a summary and outlook is given in Section\ref{set5}.

\section{Model description and observables}
\label{set0}
The UrQMD model has been widely used to study nuclear reactions within a
large range of beam energies, from the Fermi energy (tens of MeV per nucleon) up to the highest energy (TeV) presently available at the Large Hadron Collider~\cite{Bass98,Bleicher:1999xi,Li:2011zzp,Li:2012ta}. In the UrQMD model, each hadron can be represented by a Gaussian wave packet~\cite{Bass98}. Usually, the width parameter of 2 fm$^2$ is chosen for simulating collisions with Au. Mean field potential and collision terms are two of the most important ingredients of the UrQMD model. In this section, we briefly discuss the recent updates on these two terms.

\subsection{Mean field potential}
After carefully choosing nuclei with a proper binding energy and radius in the initialization, the coordinate $\textbf{r}_i$ and momentum $\textbf{p}_i$ of nucleon $i$ are propagated according to
\begin{eqnarray}
\dot{\textbf{r}}_{i}=\frac{\partial  \langle H  \rangle}{\partial\textbf{ p}_{i}},
\dot{\textbf{p}}_{i}=-\frac{\partial  \langle H \rangle}{\partial \textbf{r}_{i}}.
\end{eqnarray}
Here, {\it $\langle H \rangle$} is the total Hamiltonian function, it consists of the kinetic energy $T$ and the effective interaction potential energy $V$. For studying HICs at intermediate energies, the following density and momentum dependent potential has been widely employed in QMD-like models \cite{Aichelin:1991xy,Hartnack:1997ez,Li:2005gfa,Zhang:2018rle},
\begin{equation}\label{eq2}
V=\alpha\left(\frac{\rho}{\rho_0}\right)+\beta\left(\frac{\rho}{\rho_0}\right)^{\eta} + t_{md} \ln^2[1+a_{md}(\textbf{p}_{i}-\textbf{p}_{j})^2]\frac{\rho}{\rho_0}.
\end{equation}

Here $t_{md}$=1.57 MeV and $a_{md}$=500 $c^{2}$/GeV$^{2}$. In present version, $\alpha$, $\beta$, and $\eta$ are calculated using Skyrme parameters via $\frac{\alpha}{2}=\frac{3}{8}t_{0}\rho_{0}$, $\frac{\beta}{\eta+1}=\frac{1}{16}t_{3}\rho_{0}^{\eta}$, and $\eta=\sigma+1$. The parameters
$t_{0}$, $t_{1}$, $t_{2}$, $t_{3}$ and $x_{0}$, $x_{1}$, $x_{2}$, $x_{3}$, $\sigma$ are the well-known parameters of the Skyrme force\cite{Zhang:2006vb,Zhang:2007gd}. Following recent progress in the study of the density-dependent nuclear symmetry energy and to better describe the recent experimental data at intermediate energies, the surface, the surface asymmetry term, the symmetry energy term obtained from the Skyrme potential energy density functional have been introduced to the present version~\cite{Wang:2013wca,Wang:2014rva}. It reads as
\begin{equation}\label{urho}
\begin{aligned}
u_{Skyrme}=&u_{sur}+u_{sur,iso}+u_{sym}\\
&=\frac{g_{\text{sur}}}{2\rho_{0}}(\nabla\rho)^{2}+\frac{g_{\text{sur,iso}}}{2\rho_{0}}[\nabla(\rho_{n}-\rho_{p})]^{2}\\
&+\left(A_{\text{sym}}\frac{\rho^{2}}{\rho_{0}}+B_{\text{sym}}\frac{\rho^{\eta+1}}{\rho_{0}^{\eta}}+C_{\text{sym}}\frac{\rho^{8/3}}{\rho_{0}^{5/3}}\right)\delta^2.
\end{aligned}
\end{equation}
And, the parameters $g_{\text{sur}}$, $g_{\text{sur,iso}}$, $A_{sym}$, $B_{sym}$, and $C_{sym}$ are related to the Skyrme parameters via
\begin{eqnarray}
  \frac{g_{sur}}{2} &=& \frac{1}{64}(9t_{1}-5t_{2}-4x_{2}t_{2})\rho_{0}, \\
  \frac{g_{sur,iso}}{2} &=& -\frac{1}{64}[3t_{1}(2x_{1}+1)+t_{2}(2x_{2}+1)]\rho_{0},\\
  A_{sym} &=& -\frac{t_{0}}{4}(x_{0}+1/2)\rho_{0}, \\
  B_{sym} &=& -\frac{t_{3}}{24}(x_{3}+1/2)\rho_{0}^{\eta}, \\
  C_{sym} &=& \frac{1}{24}\left(\frac{3\pi^{2}}{2}\right)^{2/3}\rho_{0}^{5/3}\Theta_{sym},
\end{eqnarray}
where $\Theta_{sym}=3t_{1}x_{1}-t_{2}(4+5x_{2})$.
With the introduction of the Skyrme potential energy density functional, one can easily choose different Skyrme interactions to study properties of dense nuclear matter formed in HICs with the UrQMD model. E.g., to investigate the incompressibility $K_0$ of isospin symmetric nuclear matter, one can select Skyrme interactions which yield similar values of the nuclear symmetry energy but different values of $K_0$. While Skyrme interactions which give similar value of $K_0$ but very different density-dependent nuclear symmetry energy can be selected to study the effect of the nuclear symmetry energy on various observables.

The potentials for produced mesons, i.e., pion and kaon, also can be incorporated into the UrQMD model, it is found that, with considering the kaon potential (including both the scalar and vector aspects) and pion potential, the collective flow of pion and kaon can be reproduced as well, details can be found in Refs.\cite{Liu:2018xvd,Du:2018ruo}.

\subsection{The in-medium nucleon-nucleon cross section}

Besides the mean field potential, the nucleon-nucleon cross section (NNCS) is one of the most essential ingredients of the transport model as well. In free space, the information of NNCS has been well measured by experiments, but in the nuclear medium, how the NNCS varies with the nuclear density and momentum is still an open question. It is known from many theoretical studies that the NNCS in the in-medium is smaller than that in the free space, however, the degree of this reduction is still far from being entirely pinned down\cite{lgq14,Sammarruca:2005tk,HJS,CF,HFZ07,WGL,Alm:1995chb,Mao:1994zza,Li:2003vd}. One of possible way to obtain the detailed information of the in-medium NNCS is to compare the transport model simulations with the corresponding experimental data. Several different forms of the in-medium NNCS have been used in transport models, such as $\sigma_{NN}^{\text{in-medium}}=(1-\eta\rho/\rho_{0})\sigma_{NN}^{\text{free}}$ with $\eta=0.2$\cite{gd10,Zhang:2007gd}, $\sigma_{NN}^{\text{in-medium}}=0.85\rho^{-2/3}/\tanh(\frac{\sigma^{\text{free}}}{0.85\rho^{-2/3}})$\cite{dds}, $\mathcal{F}=\sigma_{NN}^{\text{in-medium}}/\sigma_{NN}^{\text{free}}=(\mu_{NN}^{*}/\mu_{NN})^2$, where $\mu_{NN}^{*}$ and $\mu_{NN}$ are the $k$-masses of the colliding nucleons in the medium and in free space \cite{Li:2005iba,Feng:2011eu,Guo:2013fka}.

\begin{figure}[htbp]
\centering
\includegraphics[angle=0,width=0.4\textwidth]{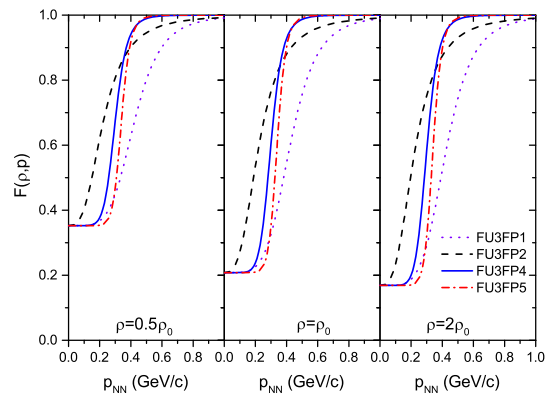}
\caption{\label{NNCS}(Color online) The medium correction factor $F(\rho,p)$ obtained with the parameterization on the momentum dependence with the four options FP1, FP2, FP4, and FP5 given in Table \ref{tabfp} at $\rho$=0.5$\rho_0$, $\rho_0$, and 2$\rho_0$.}
\end{figure}

In the present UrQMD model, the in-medium elastic NNCS are treated as the product of a medium correction factor $F$ and the cross sections in free space for which the experimental data are available. The total nucleon-nucleon binary scattering
cross sections can thus be expressed as
\begin{equation}
\sigma_{tot}^{*}=\sigma_{in}+\sigma_{el}^{*}=\sigma_{in}+F(\rho,p)
\sigma_{el}  \label{ecsf}
\end{equation}
with
\begin{equation}
F(\rho,p)=\left\{
\begin{array}{l}
f_0 \hspace{3.45cm} p_{NN}>1 {\rm GeV}/c \\
\frac{F_{\rho} -f_0}{1+(p_{NN}/p_0)^\kappa}+f_0 \hspace{1cm}
p_{NN} \leq 1 {\rm GeV}/c
\end{array}
\right.
\label{fdpup}
\end{equation}
where $p_{NN}$ denotes the momentum in the two-nucleon center-of-mass (c.o.m.) frame. Here
$\sigma_{el}$ and $\sigma_{in}$ are the NN elastic and
inelastic cross sections in free space, respectively. We note here that the experimental data of the inelastic cross
sections in free space are still be used. Because the probability for a nucleon to undergo inelastic scattering and to become a $\Delta$ is small in HICs around 1 GeV$/$nucleon regime\cite{Bass:1995pj}. Thus the influence of inelastic channels on nucleonic observables, which are mainly focused on in this work, can be neglected. The density-dependent factor $F_\rho$ is parameterized as
\begin{equation}
F_\rho=\lambda+(1-\lambda)\exp[-\frac{\rho}{\zeta\rho_0}]. \label{fr}
\end{equation}
In this work, $\zeta$=1/3 and $\lambda$=1/6 are used which correspond to FU3 in Ref.~\cite{Li:2011zzp}. To systematically investigate the effect of the in-medium NNCS on various observables in HICs at intermediate energies, four different parametrization sets for
$f_{0}$, $p_{0}$ and $\kappa$ in Eq.\ \ref{fdpup} are chosen to obtain different momentum dependences of $F(\rho,p)$. The reduced factors obtained with these parametrization sets are displayed in Fig.\ref{NNCS}. Specifically, the parametrization set FU3FP1 was usually
applied to investigate HICs around the Fermi energy region where the mean-field potential and the Pauli blocking effects are much more important. It has been found that with FU3FP1 parameter set, the experimental data of both the collective flow and the nuclear stopping power at the Fermi energy domain can be reproduced\cite{Li:2011zzp,Guo:2012aa,Wang:2012sy,lipc,Li:2018bus}. While, at higher energies, e.g., FU3FP2 has been used to extract the density-dependent symmetry energy with the elliptic flow data, as calculations with this parametrization set are found to be much more close to the experimental data of collective flows at 400 MeV$/$nucleon\cite{Russotto:2011hq}.
We further introduce the FP4 and FP5 sets which lie
roughly between FP1 and FP2. This permits more detailed studies of the
momentum dependence of the in-medium NNCS by taking advantage of the large
number of new FOPI data.

\begin{table}
\begin{center}
\renewcommand{\arraystretch}{1.2}
\begin{tabular}{|l|c|c|c|}\hline
\bf Set & $f_0$ & $p_0$ [GeV/$c$]   & $\kappa$  \\\hline\hline
\tt FP1 & 1   & 0.425 & \ 5  \        \\
\tt FP2 & 1   & 0.225 & 3        \\
\tt FP4 & 1   & 0.3 & 8        \\
\tt FP5 & 1   & 0.34 & 12        \\ \hline
\end{tabular}
\end{center}
\caption{The parameter sets FP1, FP2, FP4 and FP5 used for
describing the momentum dependence of $F(\rho,p)$. } \label{tabfp}
\end{table}

Usually, the UrQMD transport program stops at 150~fm$/c$ and an isospin-dependent minimum spanning tree (iso-MST)
method which was introduced by Zhang {\it et al.}~\cite{Zhang:2012qm} is used to construct clusters. In this method, if the relative distances and momenta of two nucleons are smaller than $R_{0}$ and $P_{0}$, respectively, they are considered to belong to the same fragment. It is found that with a proper set of $R_{0}$ and $P_{0}$, the fragment mass distribution in HICs at intermediate energies can be reproduced well \cite{Zbiri:2006ts,Russotto:2011hq,Li:2016mqd}. The parameters adopted in this paper are $R_{0}^{pp}$=2.8 fm, $R_{0}^{nn}$=$R_{0}^{np}$=3.8 fm, and $P_{0}$=0.25 Gev/c. We would like to note here that the collective flow and the nuclear stopping power, which will be focused in this work, are insensitive to $R_{0}$ and $P_{0}$, as well as the stopping time, when they are selected in their reasonable ranges \cite{Wang:2013wca}.

\subsection{Observables}

The directed $v_{1}$ and elliptic $v_{2}$ flows are the two of most widely studied observables in HICs at energy from intermediate energies to the relativistic energies, which can be obtained from the Fourier expansion of the azimuthal distribution of detected particles \cite{Reisdorf:1997fx,FOPI:2011aa,Heinz:2013th},
\begin{equation}\label{v1}
  v_{1}\equiv \langle cos(\phi)\rangle=\left\langle\frac{p_{x}}{p_{t}}\right\rangle,
\end{equation}
\begin{equation}\label{v2}
  v_{2}\equiv \langle cos(2\phi)\rangle=\left\langle\frac{p_{x}^{2}-p_{y}^{2}}{p_{t}^{2}}\right\rangle,
\end{equation}
in which $p_{x}$ and $p_{y}$ are the two
components of the transverse momentum $p_{t}=\sqrt{p_{x}^{2}+p_{y}^{2}}$. And the angle brackets in Eq.\ref{v1} and Eq.\ref{v2} indicate an average over all considered particles from all events. The directed flow $v_1$ characterizes particle motion (bounce-off or rotational-like) in the reaction plane (defined by the impact parameter $b$ in the $x$-axis and the beam direction $z$-axis), while the elliptic flow $v_2$ describes the emission (squeeze-out) perpendicular to the reaction plane. Both $v_1$ and $v_2$ have complex multi-dimensional structure. For a certain species of particles produced in a nuclear reaction with fixed colliding system, beam energy, and impact parameter, they depend both on the rapidity $y_z$ and the transverse momentum $p_t$. The scaled units $y_0\equiv y/y_{pro}$ and $u_{t0}\equiv u_t/u_{pro}$ (with $u_t=\beta_t\gamma$ the transverse component of the four-velocity and $u_{pro}$ is the velocity of the incident projectile in the c.o.m system of two nuclei) are used instead of $y_z$ and $p_t$ throughout, in the same way as done in the experimental report\cite{FOPI:2011aa}, in order to scale with whole incident energies. The subscript $pro$ denotes the incident projectile in the c.o.m system.

Usually, the slope of $v_1$ and the value of $v_2$ at mid-rapidity ($y_0$$\sim$0) are calculated and compared to the experimental data to extract the nuclear EOS and the in-medium NNCS\cite{Danie02,Ollitrault:1997vz}. Roughly speaking, at low beam energies ($\le$ 100 MeV$/$nucleon), the slope of $v_1$ is negative while the $v_2$ is positive, nucleons are more likely to be emitted in the reaction plane and undergo a rotation-like motion\cite{Andronic:2006ra}. With increasing beam energy, the slope of $v_1$ is increasing to a maximal (positive) value while the $v_2$ is decreasing to a minimal (negative) value at beam energies about 400-600 MeV$/$nucleon\cite{Andronic:2004cp,Le}. Further increasing beam energy, the slope of $v_1$ decreases while the $v_2$ increases with beam energy. In general, both $v_{1}$ and $v_{2}$, as well as nuclear stopping power in HICs around 1 GeV$/$nucleon are strongly related to the detailed ingredients of the nuclear EOS and the in-medium NNCS\cite{Zheng:1999gt,Persram:2001dg,Andronic:2004cp,Gaitanos:2004ic,Li:2005jy,Zhang:2006vb,BALi08,Li:2011zzp,Kaur:2016eaf,Barker:2016hqv,Basrak:2016cbo}.

The nuclear stopping power which measures the efficiency of converting the beam energy in the longitudinal direction into the transverse direction is also one of the most important observables. Serval different definitions$/$quantities of nuclear stopping power have been used and reported in literature, such as the quadrupole momentum tensor $Q_{zz}$=$\sum_{i}2p_z^2(i)-p_x^2(i)-p_y^2(i)$, and the ratio of transverse to parallel energy $R_E$ \cite{Lehaut:2010zz}, the ratio of the variances of the transverse rapidity distribution over that of the longitudinal rapidity distributions $varxz$ \cite{Reisdorf:2004wg}. In present work, we mainly focus on $varxz$, which reads

\begin{equation}
varxz=\frac{<y_{x}^2>}{<y_{z}^2>} . \label{eqvartl}
\end{equation}
Here
\begin{equation}
<y_{x,z}^2>=\frac{\sum(y^2_{x,z}N_{y_{x,z}})}{\sum
N_{y_{x,z}}}, \label{eqgm}
\end{equation}
where $<y_{x}^2>$ and $<y_{z}^2>$ are the
variances of the rapidity distributions of nucleons in the $x$ and
$z$ directions, respectively. $N_{y_{x}}$ and
$N_{y_{z}}$ denote the numbers of nucleons in each of the $y_x$
and $y_z$ rapidity bins. Apparently, one expects that for full stopping, the value of $varxz$ will be unity, while it will be zero for full transparency. The excitation function of the stopping power from the Fermi energy to several GeV has shown that $varxz$ first increases to a maximal value (close to but smaller than unity) at beam energy around 800 MeV$/$nucleon then decreases afterwards \cite{Andronic:2006ra,Reisdorf:2004wg,FOPI:2010aa}.
Besides the collective flow and the nuclear stopping power, other observables such as particle yield, fragment multiplicity distribution, rapidity distribution, kinetic energy and transverse momentum spectra are also applied widely to deduce the properties of the formed dense nuclear matter.

\section{Influence of the in-medium nucleon-nucleon cross section on observables}
\label{set1}

\begin{figure}[htbp]
\centering
\includegraphics[angle=0,width=0.48\textwidth]{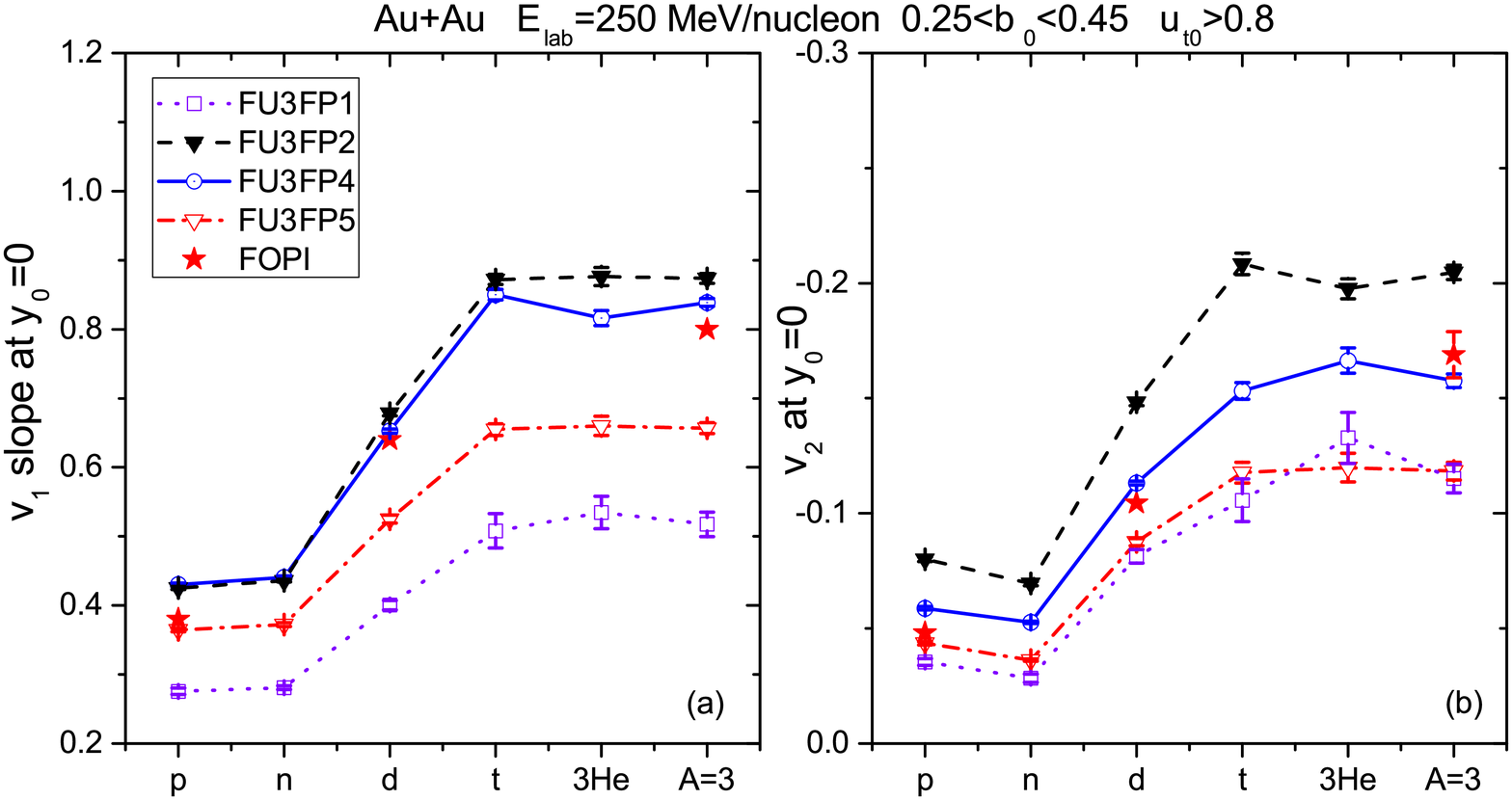}
\caption{\label{flow}(Color online) (a) The slope of the directed flow and the elliptic flow (b) at mid-rapidity ($y_0$=0) for
light particles up to mass number $A=3$ ($^3H$ and $3^He$) calculated with FU3FP1, FU3FP2, FU3FP4 and FU3FP5 (lines with symbols) parametrizations. The $^{197}$Au+$^{197}$Au collision
at $E_{\rm lab}$=250 MeV$/$nucleon with $0.25<b_0<0.45$ is considered as an example. The FOPI
experimental data (stars) are from Ref.\ \cite{FOPI:2011aa}. }
\end{figure}

\begin{figure}[htbp]
\centering
\includegraphics[angle=0,width=0.48\textwidth]{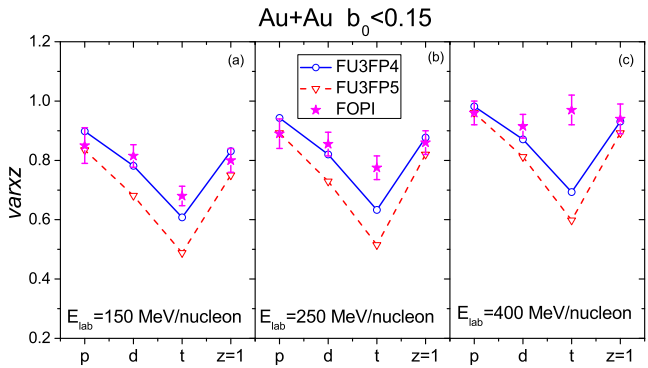}
\caption{\label{varxz}(Color online) The nuclear stopping power $varxz$ of free protons, deuterons, tritons, as well as hydrogen isotopes ($Z=1$) produced in central $^{197}$Au+$^{197}$Au collisions at the beam energies 150, 250, and 400 MeV$/$nucleon. Calculations with the FU3FP4 and FU3FP5 sets are compared with the FOPI
experimental data (stars) \cite{FOPI:2010aa}. }
\end{figure}

To show the effect of the in-medium NNCS on varies observables, $^{197}$Au+$^{197}$Au collisions at beam energies 150, 250, and 400 MeV$/$nucleon for centrality $0<b_0<0.45$ are calculated. The reduced impact parameter $b_0$ is defined as
$b_0=b/b_{max}$ with $b_{max} = 1.15 (A_{P}^{1/3} + A_{T}^{1/3})$~fm. The slope of the $v_1$ and $v_2$ at mid-rapidity for light particles are displayed in Fig.\ref{flow}. One sees clearly that
calculations with FU3FP4 (blue line) and FU3FP5 (red line) are well separated. It implies
that the directed and elliptic flows are very sensitive to the momentum dependence of the in-medium
NNCS within a narrow region of $p_{NN}=0.2-0.4$
GeV/$c$, as the largest difference between these two parametrizations exist in that narrow region (shown in Fig.\ref{NNCS}). Both the slope of $v_{1}$ calculated with FU3FP2 and FU3FP4 and
the $v_{2}$ calculated with FU3FP1 and FU3FP5
track each other closely. Large
difference between FU3FP2 and FU3FP4 at the low momenta and
between FU3FP1 and FU3FP5 at high momenta can be observed in Fig.\ref{NNCS}. Thus one may conclude that the slope of the directed flow is not sensitive to the
low momentum part while the elliptic flow is not sensitive to the high
momentum part of the in-medium NNCS. However, the sensitivity of the collective flow to the FU3FP4 and FU3FP5 sets will be reduced at higher beam energies since they almost overlap at higher relative momentum. Further, it can be seen that both the $v_1$ slope and the $v_2$ of free protons at mid-rapidity can be quite well reproduced with FU3FP5, while that of deuterons and A=3 clusters calculated with FU3FP4 are found to be more close to the experimental data than that with FU3FP5. Consequently, the FU3FP4 and FU3FP5 parametrization sets offer the greatest possible degree of the momentum-dependent in-medium NNCS. Besides the in-medium NNCS, other ingredients in transport models, such as, the initialization, the nuclear EOS, as well as the Pauli blocking effects, may also affect the collective flows and the nuclear stopping power to some extensive, details can be found in our previous publications\cite{Li:2011zzp,Wang:2013wca,lipcjpg}.

Figure.\ref{varxz} displays the nuclear stopping power $varxz$ of free protons, deuterons, tritons, as well as hydrogen isotopes ($Z=1$) in $^{197}$Au+$^{197}$Au collisions at the beam energies 150, 250, and 400 MeV$/$nucleon. Once again, it is found that the $varxz$ of free protons can be well reproduced both with FU3FP4 and FU3FP5, while the results of other light clusters calculated with FU3FP4 are found to be more close to the experimental data. The $varxz$ obtained with FU3FP5 is smaller than that with FU3FP4. Because FU3FP5 denotes a larger reduction on the NNCS at lower relative momenta than FU3FP4 does, the more violent collision prevailing in FU3FP4 parametrization enhances the nuclear stopping power.

Besides the total in-medium NNCS, the differential cross section, i.e., the angular distribution, in transport model also plays an important role. As discussed in our previous publication\cite{Wang:2016yti}, by comparing the results of the collective flows and stopping power calculated with different angular distributions within the UrQMD model, it is found that both the collective flows and the nuclear stopping power obtained by using the forward-backward peaked differential NNCS are smaller than that with the isotropic one, while the elliptic flow difference between neutrons and hydrogen isotopes can hardly be influenced by the angular distributions. Details can be found in Ref. \cite{Wang:2016yti}.

\section{Determination of the nuclear incompressibility}
\label{set2}

\begin{figure}[htbp]
\centering
\includegraphics[angle=0,width=0.48\textwidth]{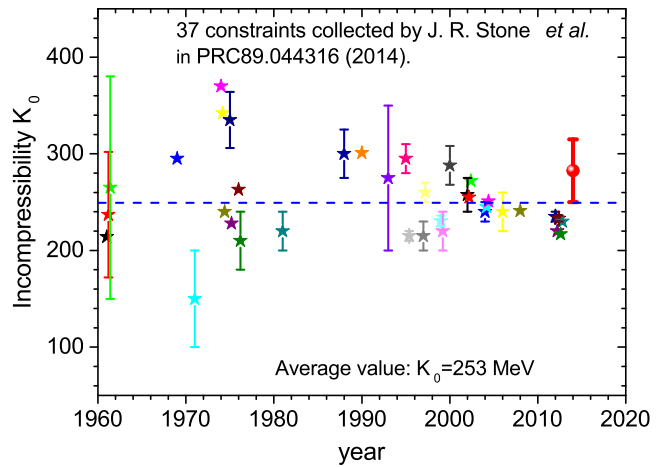}
\caption{\label{k0}(Color online) The incompressibility of isospin symmetric nuclear matter from 37 analyses using nuclei structure observations collected by J. R. Stone {\it et al.} in Ref.\cite{Stone:2014wza}.
The red circle denotes the result deduced by J. R. Stone {\it et al.} in Ref.\cite{Stone:2014wza}. The dashed line represents the averaged value.}
\end{figure}

\begin{table}[htbp]
\centering
\caption{\label{tab:table1} Saturation properties of nuclear matter as obtained with
the three Skyrme interactions used in studying the incompressibility $K_0$.}
\setlength{\tabcolsep}{1.4pt}
\begin{tabular}{lccccccc}
&$K_0$ (MeV)
&&$S_0$ (MeV)
&&$L $ (MeV)\\
\hline
Skxs15 &201 &&31.9&&34.8 \\
MSK1 &234&&30.0&&33.9\\
SKX &271&&31.1&&33.2\\
\hline

\end{tabular}

\end{table}

The EOS of isospin symmetric nuclear matter can be expanded as $\frac{E}{A}(\rho)=E_0+\frac{K_0}{18}(\frac{\rho-\rho_0}{\rho_0})^2+...$, therefore, a more accurate value of $K_0$ means a better understanding of the nuclear EOS around the normal density. Constraints on $K_{0}$ through comparing experimental data on nuclear structure
properties and theoretical model calculations have
been summarized in Ref.~\cite{Stone:2014wza}, and the results are displayed in Fig.\ref{k0}. As can be seen in Fig.\ref{k0}, most of these constraints indicate that $K_0$ should be in the range 200-300 MeV. While in Ref.\cite{Stone:2014wza}, the authors showed that $250<K_0<315$ MeV can be obtained, based on the up-to-date data on the giant monopole resonance energies. In Ref.\cite{Khan:2013mga}, the authors studied the giant monopole resonance energies of $^{208}$Pb and $^{120}$Sn, based on the constrained Hartree-Fock-Bogoliubov approach, $190<K_0<270$ MeV is found out. Although the incompressibility $K_0$ has been extensively investigated, different models offer a wide range of results for $K_0$, see, e.g., Refs. \cite{Stone:2014wza,Khan:2013mga,Giuliani:2013ppnp} and references therein.

Extraction of the incompressibility $K_0$ with HIC also has a long history, to our best knowledge, the very first studies can be found in 1980s\cite{Molitoris:1986pp,Molitoris:1985gs,Kruse:1985hy,Aichelin:1986ss,Stoecker:1986ci,Cassing:1990dr}. The collective flow and particle (e.g., $\pi$ and kaon) productions are two of the main observables used to extract $K_0$.
Using the microscopic Vlasov-Uehling-Uhlenbeck (VUU) model, evidence for a stiff ($K_0 \sim 380$ MeV) nuclear EOS was presented from a comparison with experimental data on pion production and collective sidewards flow by Joseph Molitoris and Horst St\"{o}cker et al. in 1985 \cite{Molitoris:1986pp,Molitoris:1985gs,Kruse:1985hy}. By comparing pBUU model calculations to the directed and elliptic flows in Au+Au at
the beam energies from 0.15 to 10.0 GeV$/$nucleon, the
most extreme cases (for $K_{0}$ larger than 380~MeV or less than 167~MeV)
have been ruled out by Danielewicz\emph{ et al}. \cite{Danie02}. In Refs \cite{Hartnack:2005tr,Feng:2011dp}, it was found that calculations with the soft EOS ($K_{0}$=200 MeV) are close to the kaon yields and yield ratios.

\begin{figure}[htbp]
\centering
\includegraphics[angle=0,width=0.48\textwidth]{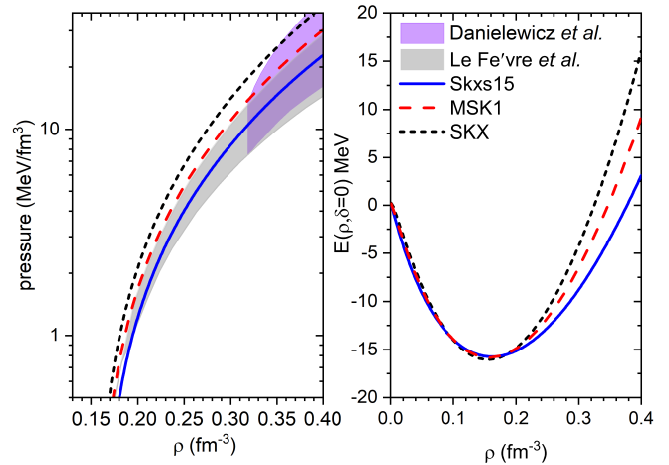}
\caption{\label{press}(Color online) The pressure and the binding energy per nucleon in symmetric nuclear matter as a function of density.
The lines represent calculations for the Skxs15, MSK1, and SKX interactions.
The results obtained by Danielewicz et al.\cite{Danie02} and Arnaud Le F\`evre et al.\cite{Fevre:2015fza} are represented by shaded regions.}
\end{figure}

Recently, $v_{2n}$ which relates to the elliptic flow ($v_2$) in a broader rapidity range has been found to be very sensitive to the incompressibility $K_0$\cite{Fevre:2015fza}. By comparing the FOPI data with the calculations using the isospin quantum molecular dynamics (IQMD) model, a incompressibility $K_0=190 \pm 30$ MeV was extracted\cite{Fevre:2015fza}. In view of the fact that the collective flow also can be influenced by the in-medium NNCS and the findings from the comparison of the transport models, i.e., results from different transport models are diversified even the same physical inputs are required~\cite{Xu:2016lue}, more studies on $v_{2n}$ seems quite necessary.

To constrain the incompressibility $K_0$ using $v_{2n}$, the Skxs15, MSK1, and SKX interactions which give quite similar values of nuclear symmetry energy but the incompressibilities $K_0$ varies from 201 MeV to 271 MeV (see Table \ref{tab:table1})\cite{Dutra:2012mb} are considered. The binding energy per nucleon and the pressure as a function of the density are illustrated in Fig.\ref{press}. For comparison, constraints obtained by Danielewicz et al.\cite{Danie02} and by F\`evre et al.\cite{Fevre:2015fza} are also displayed with shaded regions.

\begin{figure}[htbp]
\centering
\includegraphics[angle=0,width=0.48\textwidth]{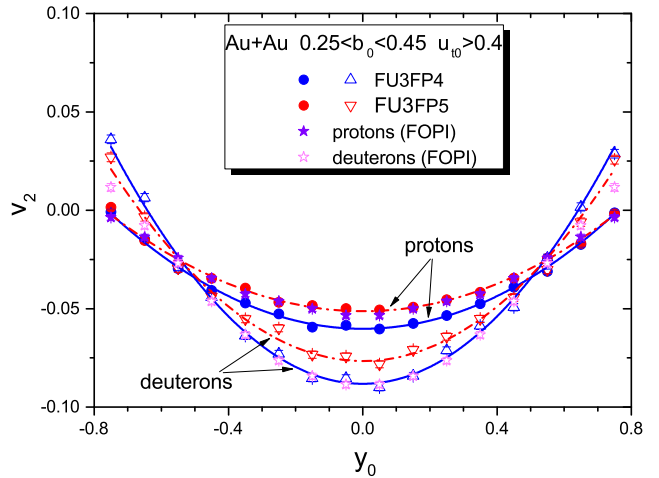}
\caption{\label{y-v2}(Color online) The elliptic flow of free protons and deuterons in Au+Au collisions at $E_{\rm lab}=0.4$~GeV$/$nucleon with centrality $0.25<b_0<0.45$ and the scaled transverse velocity $u_{t0}>0.4$. Results calculated with MSK1 together with the FU3FP4 (blue) and FU3FP5 (red) parametrizations of the in-medium NNCS are compared with the FOPI experimental data\cite{FOPI:2011aa}.
 Lines are fits to the calculated results assuming $v_2(y_0)=v_{20} + v_{22}\cdot y_0^2 $.}
\end{figure}

A good agreement between model calculations and the measured data of the elliptic flow are illustrated in Fig.\ref{y-v2} and figures in Ref. \cite{Wang:2018hsw}. Fig.~\ref{y-v2} compares the elliptic flow of free protons and deuterons calculated with FU3FP4 and FU3FP5 to the FOPI experimental data. $v_2(y_0)=v_{20} + v_{22}\cdot y_0^2 $ is used to fit the calculated results, the same as in the FOPI analysis. Both the elliptic flow of free protons and deuterons in the whole inspected rapidity range can be reproduced by calculations with FU3FP4 and FU3FP5. As it has been discussed in Refs. \cite{Fevre:2015fza} and \cite{Wang:2018hsw}, the sensitivity to the incompressibility $K_0$ is enhanced by using the observable $v_{2n}=|v_{20}| + |v_{22}|$. Because a smaller value of $K_0$ leads to smaller values of both $|v_{20}|$ and $|v_{22}|$, as can be found in figures in Refs. \cite{Fevre:2015fza,Wang:2018hsw}.

\begin{figure}[htbp]
\centering
\includegraphics[angle=0,width=0.48\textwidth]{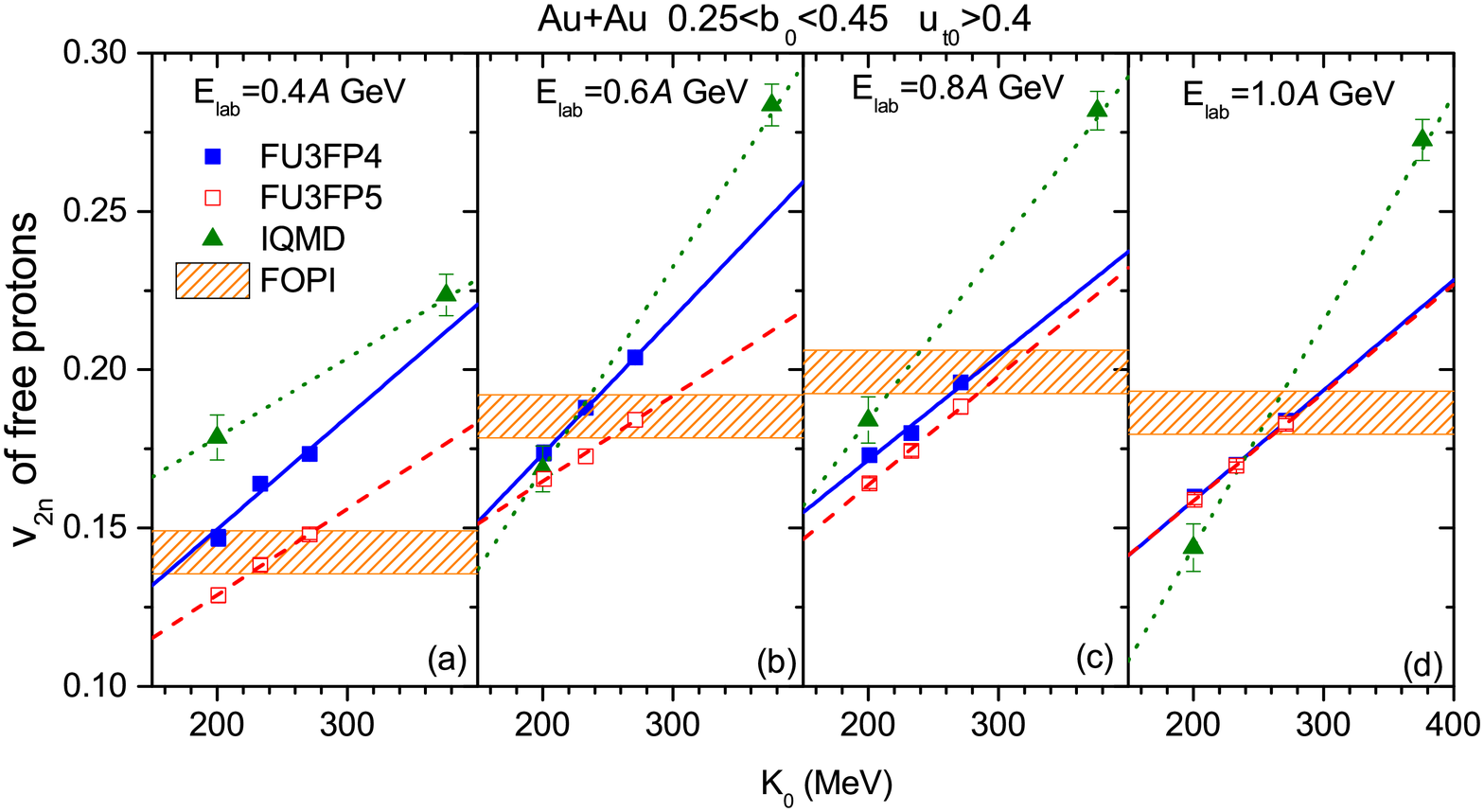}
\caption{\label{v2n-p}(Color online) The $v_{2n}$ of free protons produced from $^{197}$Au+$^{197}$Au collisions at $E_{\rm lab}=0.4$, $0.6$, $0.8$, and $1.0$~GeV$/$nucleon are shown as a function of the incompressibility $K_0$.
The results obtained from the IQMD model are represented by full triangles \cite{Fevre:2015fza}. The shaded bands indicate the FOPI experimental data. Three full squares (open squares)
denote respectively the results calculated using Skxs15, MSK1, and SKX together with the FU3FP4 (FU3FP5) parametrizations for the in-medium NNCS. The lines are the linear fits to the calculations. Reproduced from Ref. \cite{Wang:2018hsw}.}
\end{figure}

\begin{figure}[htbp]
\centering
\includegraphics[angle=0,width=0.48\textwidth]{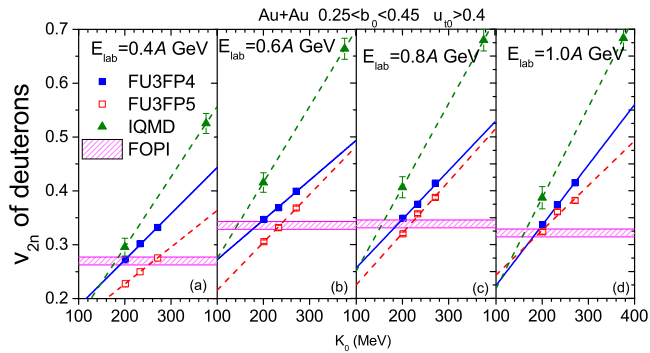}
\caption{\label{v2n-d}(Color online) The same as Fig.\ref{v2n-p} but for the $v_{2n}$ of deuterons. Reproduced from Ref. \cite{Wang:2018hsw}.}
\end{figure}

The $v_{2n}$ of free protons and deuterons are shown as a function of the incompressibility $K_0$ in Figs.\ref{v2n-p} and \ref{v2n-d}. Calculations with the FU3FP4 and FU3FP5 sets are compared to the FOPI experimental data, as well as to the results calculated with the IQMD model, taken from Ref.\cite{Fevre:2015fza}. The $v_{2n}$ increases strongly with increasing $K_0$ in both model calculations, it implies that the $v_{2n}$ is very sensitive to the incompressibility $K_0$, though the slope is not exactly the same. Generally, the values of $v_{2n}$ calculated with the UrQMD model are smaller than that with the IQMD model, and the difference become smaller at higher beam energies. Consequently, the extracted $K_0$ with the IQMD model is smaller than that with the UrQMD model. As we have discussed in Ref. \cite{Wang:2018hsw}, the difference may comes from the different collision term in the two models, i.e., the free NNCS is used in the IQMD model, while the UrQMD model incorporates the in-medium NNCS (density- and momentum- dependent). At higher energies, the difference between the two models become smaller, it is because that the in-medium and free cross sections at the higher relative momenta are almost the same, as shown in Fig.\ref{NNCS}. Besides, different values of the width of the Gaussian wave packet and different treatments in the Pauli blocking in the two models may also contribute the observed difference in the extraction of $K_0$. Influences of these treatments and/or parameters on the $v_{2n}$ deserve further studies. On average, the central value of the incompressibility $K_0$ is obtained to be 240 (275) MeV for calculations with the FU3FP4 (FU3FP5) parametrization. With a stronger reduction of the in-medium nucleon-nucleon cross section, i.e., FU3FP5, a larger $K_0$ is extracted. It may also explain the reason why the $K_0$ obtained with the UrQMD model is larger than that with the IQMD model. $K_0 = 240 \pm 20$~MeV ($K_0 = 275 \pm 25$~MeV) for the FU3FP4 (FU3FP5) parametrization of the in-medium NNCS, which best describes the $v_{2n}$ of free protons, is extracted. In addition, within both models, it is found that $K_0$ extracted from the $v_{2n}$ of deuterons is smaller than that from $v_{2n}$ of free protons. Furthermore, the extracted $K_0$ from the $v_{2n}$ of deuterons is not sensitive to the beam energy, which is unlikely to that observed for the $v_{2n}$ of free protons. $K_0 = 190 \pm 10$~MeV ($K_0 = 225 \pm 20$~MeV) for the FU3FP4 (FU3FP5) parametrization is obtained from the $v_{2n}$ of deuterons. By combining the error intervals of the results obtained from the $v_{2n}$ of free protons and deuterons, an averaged $K_0 = 220 \pm 40$~MeV is obtained for the FU3FP4 parametrization.

\section{Constraints on the density-dependent nuclear symmetry energy}
\label{set3}

\begin{figure}[htbp]
\centering
\includegraphics[angle=0,width=0.48\textwidth]{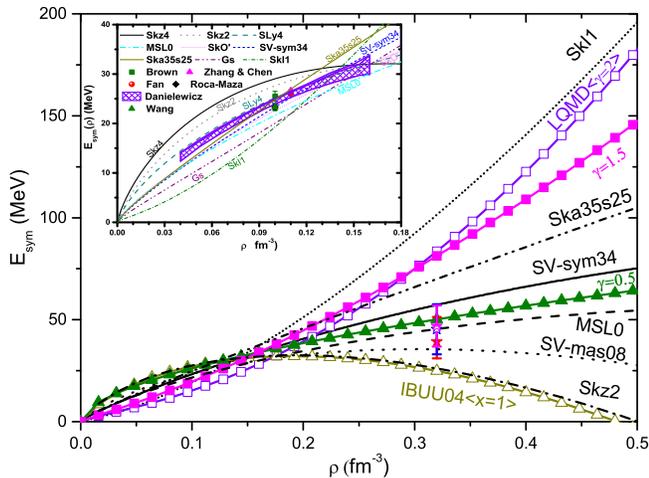}
\caption{\label{esym}(Color online) The nuclear symmetry energy
for various Skyrme interactions are displayed as a function of density. Symmetry energies used in Ref.~\cite{Russotto:2011hq} with $\gamma$=0.5 and 1.5, and favored in LQMD model~\cite{Feng:2009am} and in IBUU04 model~\cite{Xiao:2009zza} are also shown for comparison. Stars are constraints at 2$\rho_0$ obtained from astrophysical observations by Zhang and Li \cite{Zhang:2018bwq}, Xie and Li \cite{Xie:2019sqb}, and Zhou and Chen \cite{zhouchen}. The nuclear symmetry energy for Skz4, Skz2, SLy4, MSL0, SkO', SV-sym34, Ska35s25, Gs, and SkI1 at lower densities are displayed in the inset. The shaded region exhibits the result obtained
by Danielewicz \emph{et al.} \cite{Danielewicz:2013upa}.
Five different scattered symbols represent recent constraints obtained by Roca-Maza \emph{et al.} \cite{RocaMaza:2012mh}, Brown\cite{brown},
Zhang \emph{et al.} \cite{Zhang:2013wna}, Wang \emph{et al.}\cite{Wang:2015kof}, and Fan \emph{et al.} \cite{Fan:2014rha}, respectively. }
\end{figure}

To probe the density-dependent nuclear symmetry energy with HICs, microscopic transport models which provides a bridge between experimental observables and the nuclear symmetry energy are necessary. Many observables have been predicted as sensitive probes for the nuclear symmetry energy, e.g., the yield ratio and the collective flow difference (ratio) between different isospin partners (e.g., proton and neutron, $^3$H$/$$^3$He, $\pi^{-}$/$\pi^{+}$, $K^0/K^+$, and $\Sigma^{-}$/$\Sigma^{+}$), as well as the balance energy of directed flow~\cite{Li:2002qx,DiToro:2010ku,
Tsang:2008fd,Lopez:2007,Xiao:2009zza,Feng:2009am,Russotto:2011hq,Cozma:2011nr,Xie:2013np,Cozma:2013sja,
Li:2005zza,LI:2005zi,Kumar:2011td,Gautam:2010da,Lu:2016htm,Russotto:2013fza,Wang:2012sy,Guo:2012aa,Wang:2014aba,Tsang:2016foy}. In spite of the progress made, a precise constraint on the density-dependent nuclear symmetry energy with HICs is still very difficult to achieve due to a) the difficulties in precision experimental measurements, and b) strong model- and observable-dependent results, see, e.g., Refs.~\cite{Tsang:2012se,Li:2012mw,Chen:2012pk,Wolter,Li:2013ola} for review.

In this section, we review in detail two of our recent studies on the density-dependent nuclear symmetry energy by using the $^3$H$/$$^3$He yield ratio and the elliptic flow ratio between neutrons and hydrogen isotopes. As these two observables probe the nuclear symmetry energy at different density region, we incorporate two groups of Skyrme interactions into the UrQMD model. Group I includes 13 Skyrme interactions for which give quite similar values of the incompressibility $K_0$ but different values of $L$ \cite{Dutra:2012mb}, the saturation properties of these interactions are shown in Table~\ref{GroupI}. In addition, the slope parameter $L$ at $\rho$=0.08, 0.055, and 0.03 $fm^{-3}$ are also shown in Table~\ref{GroupI}. Group II includes 21 Skyrme interactions. The saturation properties of these Skyrme interactions are shown in Table~\ref{GroupII}. Moreover, the SkA and SkI5 which give larger values of the incompressibility $K_0$ are also considered to examine the influence of $K_0$ on the elliptic flow ratio and difference. The density-dependent nuclear symmetry energy from various Skyrme interactions are displayed in Fig.\ref{esym}. It can be seen that the selected Skyrme interactions cover the different forms of symmetry
energies currently discussed by different theoretical groups. In addition, some recent constraints extracted from nuclear structure properties, e.g., binding energy, neutron skin thickness, and isovector giant quadrupole resonance~\cite{RocaMaza:2012mh,Zhang:2013wna,brown,Fan:2014rha,Wang:2015kof}, and from astrophysical observations \cite{Zhang:2018bwq,Xie:2019sqb,zhouchen} are also displayed for comparison (scatter markers and shaded band).

\subsection{Result from $^3$H/$^3$He yield ratio}

\begin{table}[htbp]
\caption{\label{GroupI} Group I: Saturation properties of nuclear matter as obtained with the selected 13 Skyrme interactions used to study the $^3$H$/$$^3$He yield ratio. All entries are in MeV, except for density in $fm^{-3}$.}
\begin{tabular}{cccccccccccccc}
\hline
\hline
&
&&
&&
&&$\rho=\rho_0$
&&$\rho=0.01$
&&$\rho=0.08$
\\
&$\rho_{0}$ &&$K_0$ &&$S_0$ &&$L $  &&$L(\rho)$ &&$L(\rho)$\\
\hline
Skz4 	&	0.16	&&	230 	&&	32.0 	&&	5.8 	&&	16.5 	&&	34.5 	\\
Skz2 	&	0.16	&&	230 	&&	32.0 	&&	16.8 	&&	14.2 	&&	35.7 	\\
SV-mas08	&	0.16	&&	233 	&&	30.0 	&&	40.2 	&&	10.6 	&&	32.8 	\\
SLy4	&	0.16	&&	230 	&&	32.0 	&&	45.9 	&&	12.1 	&&	33.2 	\\
MSL0   	&	0.16	&&	230 	&&	30.0 	&&	60.0 	&&	8.7 	&&	31.6 	\\
SkO'	&	0.16	&&	222 	&&	32.0 	&&	68.9 	&&	8.9 	&&	33.2 	\\
SV-sym34 	&	0.159	&&	234 	&&	34.0 	&&	81.0 	&&	8.4 	&&	35.7 	\\
Rs	&	0.158	&&	237 	&&	30.8 	&&	86.4 	&&	6.7 	&&	31.4 	\\
Gs	&	0.158	&&	237 	&&	31.1 	&&	93.3 	&&	8.6 	&&	38.1 	\\
Ska35s25 	&	0.158	&&	241 	&&	37.0 	&&	98.9 	&&	6.3 	&&	31.7 	\\
SkI2  	&	0.158	&&	241 	&&	33.4 	&&	104.3 	&&	7.0 	&&	32.4 	\\
SkI5 	&	0.156	&&	256 	&&	36.6 	&&	129.3 	&&	6.9 	&&	34.5 	\\
SkI1 	&	0.16	&&	243 	&&	37.5 	&&	161.1 	&&	3.6 	&&	33.4 	\\

\hline
\hline
\end{tabular}
\label{skyrme}
\end{table}

\begin{figure}[htbp]
\centering
\includegraphics[angle=0,width=0.48\textwidth]{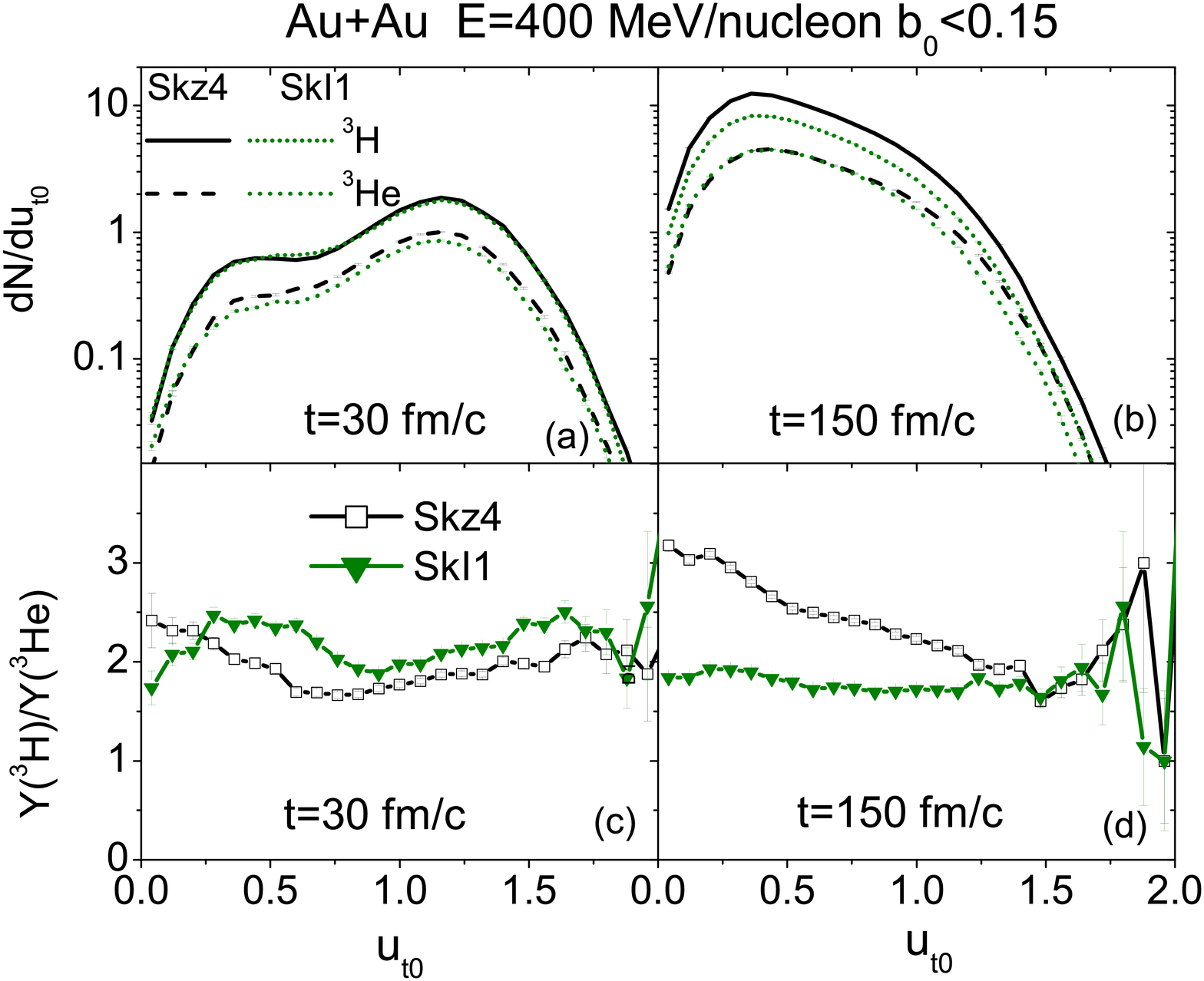}
\caption{\label{ut0}(Color online) The transverse component of the four-velocity distributions of $^3$H and $^3$He as well as the corresponding ratio at reaction times t=30 fm/c and 150 fm/c. The results calculated with two extreme cases (Skz4 and SkI1) are displayed.}
\end{figure}

\begin{figure}[htbp]
\centering
\includegraphics[angle=0,width=0.48\textwidth]{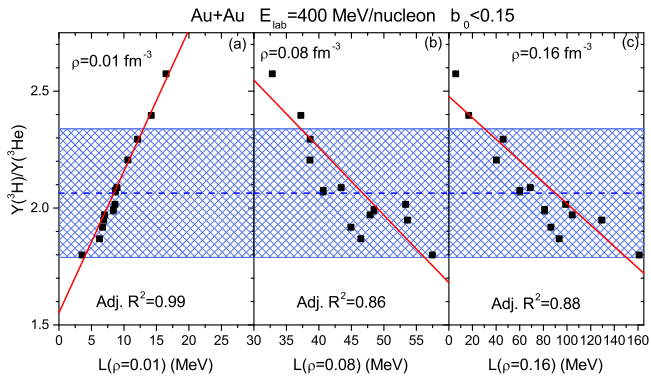}
\caption{\label{3h3he-e400}(Color online) $^3$H/$^3$He ratio as a function of the slope of $E_{sym}(\rho)$ at densities of $\rho$=0.01 $fm^{-3}$ and 0.08$fm^{-3}$, as well as the saturation density $\rho_0$. The lines represent linear fits to calculations. Correspondingly, the Adj. $R^2$ values are also given. The shaded region indicates the FOPI data of $^3$H/$^3$He ratio \cite{FOPI:2010aa}. }
\end{figure}

Calculations with both the quantum molecular dynamics (QMD) type and the Boltzmann-Uehling-Uhlenbeck (BUU) type transport models have been shown that the yield ratio of $^3$H and $^3$He emitted from HICs can be used to constrain the nuclear symmetry energy\cite{Chen:2003qj,Chen:2004kj,Li:2005kqa,Zhang:2005sm}, but some puzzling inconsistency still exists.
For example, the yield of $^3$H calculated with a soft symmetry energy is larger than that
with a stiff one based on calculations with two different QMD type models~\cite{Li:2005kqa,Zhang:2005sm}, while the opposite trend is found in Ref.~\cite{Chen:2003qj} with the isospin-dependent BUU (IBUU) model. Recently, a large amount of yield data for protons, $^2$H, $^3$H, $^3$He, and
$^4$He produced in HICs at intermediate energies has been measured by the FOPI collaboration\cite{FOPI:2010aa,FOPI:2011aa}. This data set offers new opportunities for studying the nuclear symmetry energy by using the $^3$H$/$$^3$He ratio over wide ranges of both beam energy and system size.

The transverse component of the four-velocity $u_{t0}$ distributions of $^3$H and $^3$He as well as the corresponding ratios at reaction times $t$=30 fm$/$c and 150 fm$/$c are shown in Fig.\ref{ut0}. It can be seen that, at $t$=30 fm$/$c (the early stage of expansion phase), $^3$H and $^3$He with high $u_{t0}$ are more abundant than that with low $u_{t0}$, these clusters mainly reflect the behavior of symmetry energy at high densities. At early stage, $^3$H and $^3$He consist of protons and neutrons which emitted mainly from the high density region. A stiff (i.e, SkI1) symmetry energy will repel more neutrons and less protons than a soft one (i.e., Skz4). Thus more $^3$H (neutron-rich) can be formed, then higher values of the $^3$H/$^3$He and neutron/proton ratios are obtained. As the reaction proceeds, i.e., at $t$=150 fm$/$c, more and more $^3$H and $^3$He clusters with low-$u_{t0}$ are emitted from low density environment, and finally the ratio reflects the behavior of symmetry energy at sub-saturation densities. Very similar results can be found in calculations with other QMD-type models \cite{Li:2005kqa,Kumar:2011td}. Furthermore, with increasing $u_{t0}$, the ratio calculated with Skz4 approach that of SkI1, their order may even be reversed if the residual symmetry potential is large enough. We have checked that the reversed order on the neutron$/$proton ratio is more obvious than that on the $^3$H$/$$^3$He ratio, as one excepted.

Figure~\ref{3h3he-e400} shows the $^3$H/$^3$He ratios calculated with the 13 selected Skyrme interactions as a function of the slope of $E_{sym}(\rho)$ at three different densities. The line in each bunch represents a linear fit to the calculations, the respective value of the adjusted coefficient of determination (Adj. $R^2$) is also shown. A very strong linearity between the $^3$H/$^3$He ratio and the slope of $E_{sym}(\rho)$ at $\rho$=0.01 $fm^{-3}$
can be observed, which indicates again a strong correlation between them at low densities. The results obtained with Skz4 and Skz2 fall outside the band, while it obtained with MSL0, SkO', SV-sym34, and Ska35s25 are centered in the experimental band. The symmetry energy obtained with these four interactions also lie quite close to the constraints obtained from other methods, as shown in Fig.\ref{esym}. Obviously, the large uncertainty of the experimental data prevents us from getting a tighter constraint on the density-dependent symmetry energy. However, the comparison to experimental $^3$H/$^3$He data as functions of beam energy and system size is possible, supplying a more systematic and thus more consistent information on the symmetry energy. As shown and discussed in our previous publication \cite{Wang:2014aba}, the $^3$H/$^3$He data from different collision systems (i.e.,  $^{40}$Ca+$^{40}$Ca, $^{96}$Ru+$^{96}$Ru, $^{96}$Zr+$^{96}$Zr) and different beam energies (from 0.09 to 1.5 GeV$/$nucleon) also can be well reproduced by the calculations with MSL0, SkO', SV-sym34, and Ska35s25. Although a tighter constraint on the density-dependent nuclear symmetry energy is still not obtained from the data of $^3$H/$^3$He yield ratio, partly due to the large uncertainties in the experimental data, a very satisfactory consistency among the presented comparisons is achieved. Furthermore, the results obtained from $^3$H/$^3$He yield ratio is also in agreement with our previous results obtained from the elliptic flow ratio \cite{Wang:2014rva,Russotto:2011hq,Cozma:2013sja},  although one should keep in mind that the nuclear symmetry energies at different density regions are extracted from these two observables.

\subsection{Result from the elliptic flow ratio between neutrons and hydrogen isotopes}

Using the neutron-proton differential transverse and elliptic flows to probe the isospin-dependent EOS has been proposed almost twenty year ago\cite{Li:2000bj,Li:2002qx,Greco:2002sp}, while the first constraint on the $E_{sym}(\rho)$ by using elliptic flows ratio between neutrons and hydrogen isotopes was achieved ten years ago\cite{Russotto:2011hq,Trautmann:2009kq}. By comparing the simulations of the UrQMD model and the FOPI/LAND data for Au+Au collisions at 400 MeV$/$nucleon, a moderately soft symmetry energy with a slope of $L$=83$\pm$26 MeV was obtained \cite{Russotto:2011hq}. Afterwards, new FOPI/LAND experimental data of the elliptic flow ratio between neutrons and all charged particles became available \cite{Russotto:2016ucm}, the UrQMD model also has been updated \cite{Wang:2013wca}. In this section, we review the results from the UrQMD model in which the Skyrme energy density functional is introduced to obtain parameters in the mean-field potential term.

\begin{table}[htbp]
\caption{\label{GroupII} Group II: Saturation properties of nuclear matter as obtained with the 21 Skyrme parameterizations used to study the elliptic flow ratio between neutrons and hydrogen isotopes. All entries are in MeV, except for density in $fm^{-3}$.}
\begin{ruledtabular}
\begin{tabular}{lccccccccccc}
&$\rho_{0}$
&&$K_0$
&&$S_0$
&&$L $
&&$K_{\rm sym}$ \\
\hline
Skz4 &0.160&&230 &&32.0&&5.8&&-240.9 \\
BSk8	&	0.159	&&	230 	&&	28.0 	&&	14.9&&-220.9\\
Skz2 &0.160 &&230&&32.0&&16.8&&-259.7\\
BSk5	&	0.157	&&	237 	&&	28.7 	&&	21.4 	&&-240.3\\
SkT6	&	0.161	&&	236 	&&	30.0 	&&	30.9 	&&-211.5\\
SV-kap00	&	0.16	&&	233 	&&	30.0 	&&	39.4 	&&-161.8\\
SV-mas08&0.160 &&233&&30.0&&40.2&&-172.4\\
SLy230a	&	0.16	&&	230 	&&	32.0 	&&	44.3 	&&-98.2\\
SLy5	&	0.16	&&	230 	&&	32.0 	&&	48.2 	&&-112.8\\
SV-mas07	&	0.16	&&	234 	&&	30.0 	&&	52.2 	&&-98.8\\
SV-sym32	&	0.159	&&	234 	&&	32.0 	&&	57.1 	&&-148.8\\
MSL0    &0.160&&230&&30.0&&60.0&&-99.3\\
SkO' &0.16 &&222&&32.0&&68.9&&-78.8\\
Sefm081&0.161&&237&&30.8&&79.4&&-39.5\\
SV-sym34 &0.159&&234&&34.0&&81.0&&-79.1\\
Rs&0.158&&237&&30.8&&86.4&&-9.2\\
Sefm074&0.16&&240&&33.4&&88.7&&-33.1\\
Ska35s25 &0.158&&241&&37.0&&98.9&&-23.6\\
SkI1 &0.160&&243&&37.5&&161.1&&234.7\\
\hline
SkA &0.155&&263&&32.9&&74.6&&-78.5\\
SkI5 &0.156&&256&&36.6&&129.3&&159.6\\

\end{tabular}
\end{ruledtabular}\label{skyrme}
\end{table}

\begin{figure}[htbp]
\centering
\includegraphics[angle=0,width=0.48\textwidth]{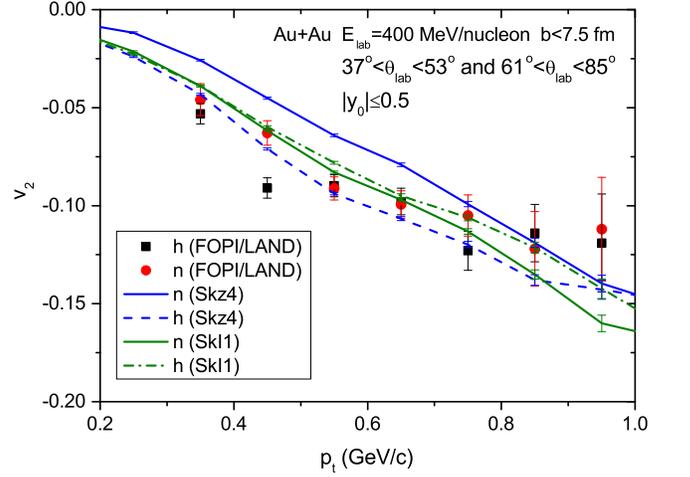}
\caption{\label{pt-v2}(Color online) Elliptic flow $v_2$ of neutrons and hydrogen isotopes for Au + Au with b $\le$ 7.5 fm and $E_{\rm lab}=0.4$~GeV$/$nucleon as a function of the transverse momentum $p_t$. The rapidity window $|y_0|<0.5$ is chosen the same as the experimental data.
 The results calculated with two extreme cases (i.e., Skz4 and SkI1) are compared to the FOPI/LAND data reported in Ref.~\cite{Russotto:2011hq}.}
\end{figure}

A good agreement between the UrQMD calculations and the experimental data can be observed again in Fig.\ref{pt-v2} where the $v_2$ of neutrons and hydrogen isotopes as a function of the transverse momentum $p_t$ is displayed. The $v_2$ of neutrons obtained with SkI1 is more negative than that obtained with Skz4, while the opposite trend is observed for hydrogen isotopes. This finding has been reported and discussed widely in Refs.\cite{Russotto:2011hq,Cozma:2013sja,Wang:2014rva,Cozma:2017bre,Trautmann:2009kq,Trautmann:2010at}. It is due to the fact that the nuclear symmetry potential tends to attract protons and expel neutrons in a neutron-rich environment, and the repulsion for neutrons (attraction for protons) are much stronger for the stiff symmetry energy (i.e., SkI1) at densities above $\rho_0$ than that for the soft one (i.e., Skz4). The stronger repulsion interaction results in the more negative elliptic flow at the beam energy studied here.

\begin{figure}[htbp]
\centering
\includegraphics[angle=0,width=0.48\textwidth]{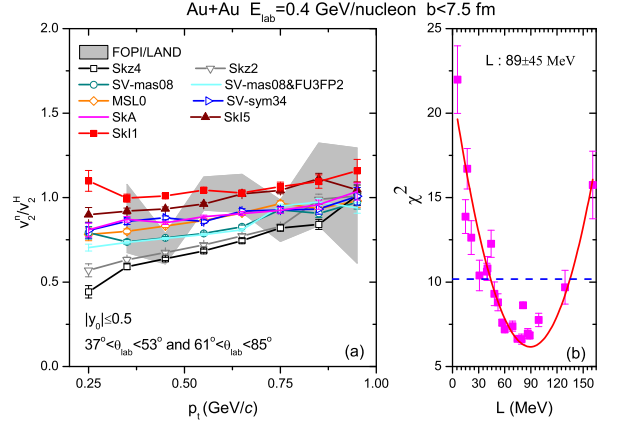}
\caption{\label{pt-v2-ratio}(Color online) (a) Elliptic flow ratio between neutrons and hydrogen isotopes $v_{2}^{n}$/$v_{2}^{H}$ as a function of the transverse
momentum $p_t$. Calculations with the indicated 9 Skyrme interactions are compared to the FOPI/LAND data (shaded area) reported in
Ref.~\protect\cite{Russotto:2011hq}. (b) The total $\chi^2$ which demonstrates the quality of the fitting procedure is plotted as a function of the slope parameter $L$. The smooth curve is a quadratic
fit to the total $\chi^2$, and the horizontal dashed line is used to determine the error of $L$
within a 2-$\sigma$ uncertainty. Reproduced from Ref.\cite{Wang:2014rva}.  }
\end{figure}

\begin{figure}[htbp]
\centering
\includegraphics[angle=0,width=0.48\textwidth]{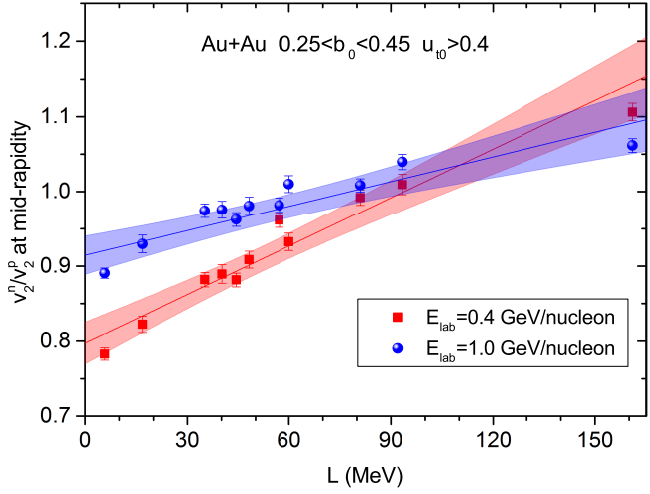}
\caption{\label{l-v2np}(Color online) Elliptic flow ratio at mid-rapidity between neutrons and protons $v_{2}^{n}$/$v_{2}^{p}$ as a function of the slope of the nuclear symmetry energy $L$. Au+Au collisions at $E_{\rm lab}=0.4$ and 1.0~GeV$/$nucleon are displayed. Calculations with the 11 Skyrme interactions listed in Ref.\cite{wyj-plb2} are shown by solid symbols. The solid lines represent linear
fits to the calculations, shaded bands are 95\% confidence intervals around the fitted lines.  }
\end{figure}

Figure~\ref{pt-v2-ratio} (a) shows the comparison of the measured and the calculated
ratios $v_{2}^{n}$/$v_{2}^{H}$ as a function of the transverse momentum $p_t$ ($p_t = u_{t0} \cdot 0.431$~GeV/c at $E_{\rm lab}=400$~MeV/nucleon for nucleons). SV-mas08\&FU3FP2 denotes the ratio calculated with SV-mas08 interaction and
the FU3FP2 parameterization of the in-medium NNCS, while others are calculated with the FU3FP4 parameterization on the in-medium NNCS. It can be seen that the $v_{2}^{n}$/$v_{2}^{H}$ ratio increases with
increasing $L$, and the difference among calculations steadily grows when moving to the low
transverse momentum region. The
results calculated with SV-sym34 and SkA (give similar value of $L$) are almost overlapped even though the difference in $K_0$ is as large as almost 30 MeV. It illustrates that the elliptic
flow ratio is not sensitive to the incompressibility $K_0$. We note here that the elliptic flows of both neutrons and hydrogen isotopes obtained with SkA are larger than that obtained with other interactions, however by taking the ratio, the impact of the incompressibility $K_0$ can be largely canceled out, similar results also can be found in Ref. \cite{Cozma:2017bre} by using T\"{u}bingen QMD model. Furthermore,
the ratio obtained with SV-mas08\&FU3FP2 lies close to that obtained with SV-mas08 in which the FU3FP4 parameterization on the in-medium NNCS is used, indicating that the influence of the in-medium NNCS on the elliptic flow ratio is quite small, similar result also has been observed in Ref.~\cite{Russotto:2011hq}. Thus, one can conclude that the systematically increasing of $v_{2}^{n}$/$v_{2}^{H}$ as displayed in Fig.\ref{pt-v2-ratio} (a) is mainly caused by the increase of the stiffness of the nuclear symmetry
energy and not caused by other changes of the isoscalar components of the mean-field potential. Fig.~\ref{pt-v2-ratio} (b) shows the quality of the fitting to the FOPI/LAND data. The total $\chi^2$ as calculated with the 21 Skyrme interactions are displayed as a function of the slope parameter $L$. It can be seen that the variation of $\chi^2$ with $L$ can be well described with a quadratic fit. The slope parameter is extracted to be $L=89\pm45$~MeV within a 2-$\sigma$ uncertainty. In Ref.\cite{Wang:2014rva}, the four
observables $v_2^{n}-v_2^{p}$, $v_2^{n}-v_2^{H}$, $v_2^{n}/v_2^{p}$, and $v_2^{n}/v_2^{H}$ (the $p_t$-integrated results) are displayed as a function of the slope parameter $L$ of the 21 Skyrme interactions. Fairly good linearities between these observables and the slope parameter are observed. Together with the FOPI/LAND data, constraints on the slope parameter $L$ can be achieved. The intervals of $L$=61-137, 44-103, 62-132, and 54-106 MeV are obtained from $v_2^{n}-v_2^{p}$, $v_2^{n}-v_2^{H}$, $v_2^{n}/v_2^{p}$, and $v_2^{n}/v_2^{H}$, respectively. Although these results are largely overlapped with each other, the largest difference among their central values is about 25 MeV (from $v_2^{n}-v_2^{p}$ and $v_2^{n}-v_2^{H}$). The uncertainties of $L$ obtained from $v_2^{n}/v_2^{p}$ and $v_2^{n}/v_2^{H}$ are smaller than that from $v^n_2$-$v^H_2$ and $v^n_2$-$v^p_2$, one of the possible reason is that by taking the ratio the impact from uncertainties in the determination of the reaction plane and in the isoscalar components of the nuclear potential can be largely cancelled out. To our knowledge, the uncertainty of the extracted $L$ using the elliptic flow ratio (difference) is large for two main reasons. (a) The large uncertainty in neutron flow measurements. (b) Contribution of $K_{sym}$ to the elliptic flow ratio has not been disentangled. As $K_{sym}$ becomes more and more important for studying the high-density behavior of nuclear symmetry energy, both $L$ and $K_{sym}$ are expected to affect the elliptic flow ratio. Thus, correlation analyses and more systematic and accurate experimental data of the elliptic flow difference (ratio) are required before achieving a tighter constraint on the density-dependent nuclear symmetry energy. Fig.\ref{l-v2np} shows the elliptic flow ratio $v_{2}^{n}$/$v_{2}^{p}$ as a function of the slope parameter $L$ at beam energies of 0.4 and 1.0~GeV$/$nucleon. With increasing beam energy, the sensitivity of $v_{2}^{n}$/$v_{2}^{p}$ to $L$ is reduced, because of the weakened mean field potential effects at higher energies. Moreover, the inelastic collisions (e.g., $n$+$n$ $\rightarrow$ $p$ + $\Delta^-$ $\rightarrow$ $p$ + $n$ +$\pi^-$) may also further reduce the effects of the symmetry potential on the flow of nucleons, as neutrons (protons) can be converted to protons (neutrons).

\begin{figure}[htbp]
\centering
\includegraphics[angle=0,width=0.48\textwidth]{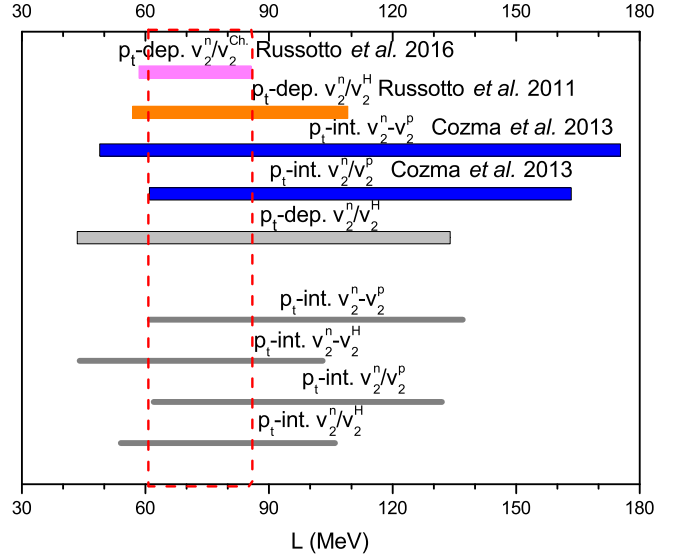}
\caption{\label{L-com}(Color online) Constraints on the slope of $E_{sym}(\rho)$ by using the elliptic flow ratio (difference) between neutrons and protons (hydrogen isotopes, all charged particles).
The results obtained by Russotto \emph{et al.} in Refs.\cite{Russotto:2011hq,Russotto:2016ucm} and by Cozma \emph{et al.} in Ref.~\cite{Cozma:2013sja} are compared to the results obtained in present work.}
\end{figure}

Constraints on the slope of $E_{sym}(\rho)$ by using the elliptic flow ratio (difference) between neutrons and protons (hydrogen isotopes, all charged particles) are summarized in Fig.\ref{L-com}. It is interesting to observe that constraints presented with the updated UrQMD model, and obtained by Cozma \emph{et al.} with the T\"{u}bingen QMD model \cite{Cozma:2013sja}, as well as obtained by Russotto \emph{et al.} with previous version of the UrQMD model \cite{Russotto:2011hq,Russotto:2016ucm}, are well overlapped within the range of 60-85 MeV. Moreover, in a recent study\cite{Cozma:2017bre}, by considering much more theoretical uncertainties in the T\"{u}bingen QMD model, e.g., the compressibility $K_0$, the in-medium nucleon-nucleon cross section, and the nucleon effective mass splitting, the slope of $E_{sym}(\rho)$ is extracted to be $L$=84$\pm$30 (exp) $\pm$ 19 (theor) MeV from the elliptic flow ratio between neutrons and protons (hydrogen isotopes). Again, this result also overlaps with $L$=60-85 MeV. Furthermore, it is noticed that the extracted central values of $L$ from $v_2^{n}-v_2^{H}$ ($v_2^{n}/v_2^{H}$) are smaller than that from $v_2^{n}-v_2^{p}$ ($v_2^{n}/v_2^{p}$). The result obtained with from the elliptic flow ratio between neutrons and all charged particles ($v_2^{n}/v_2^{Ch.}$) is also smaller than that with $v_2^{n}-v_2^{H}$ and $v_2^{n}/v_2^{H}$. It is known from the analysis in Ref.\cite{Russotto:2016ucm} that, the sensitivity densities obtained from $v_2^{n}/v_2^{H}$ and $v_2^{n}/v_2^{Ch.}$ are smaller than that with $v_2^{n}/v_2^{p}$. The difference in the extracted $L$ may also stem from the fact that different range of densities are probed by these different observables. Very recently, a new observable related to the rapidity at which $v_2$ of protons changes sign from negative to positive, is found to be sensitive to the density-dependent nuclear symmetry energy, by comparing the FOPI data and the UrQMD calculations, the slope parameter $L=43 \pm 20$ MeV is extracted \cite{wyj-plb2}. This result is about 30 MeV smaller than the result summarized in Fig.\ref{L-com}, the main reason is that this new observable which involves $v_2$ in a broader rapidity range probes the nuclear symmetry energy at a lower density region. Detailed discussions can be found in Ref. \cite{wyj-plb2}.


\section{Summary and outlook}
\label{set5}

We review our recently studies on the nuclear equation-of-state and the in-medium NNCS by using the ultrarelativistic quantum molecular dynamics (UrQMD) model. With incorporating the Skyrme potential energy density functional to obtain parameters in the mean-field potential part of the UrQMD model, three Skyrme interactions which give quite similar values of the nuclear symmetry energy but different values of the incompressibilities $K_0$ are adopted. It is found that the nuclear incompressibility $K_0$ is quite sensitive to the $v_{2n}$. By comparing the FOPI data of the $v_{2n}$ of free protons and deuterons with the UrQMD model calculations, an averaged $K_0 = 220 \pm 40$~MeV is extracted with the FU3FP4 parametrization on the in-medium NNCS. However, remaining systematic uncertainties, partly related to the choice of in-medium NNCS, are of the same magnitude ($\pm 40$~MeV). Overall, the rapidity dependent elliptic flow supports a soft nuclear equation-of-state.

With considering different forms of the density- and momentum-dependent in-medium NNCS in the UrQMD model, their influence on the collective flow and nuclear stopping power is studied as well. It is found that both the collective flow and the nuclear stopping power of free protons can be reproduced with the calculations using the FU3FP5 parametrization on the in-medium NNCS, while the results of light clusters are found to be reproduced well with the FU3FP4 parametrization. The FU3FP4 and FU3FP5 parametrization sets offer the greatest possible degree of the momentum-dependent in-medium NNCS.

 We further review the extraction of the density-dependent nuclear symmetry energy by using the experimental data of $^3$H/$^3$He yield ratio and the elliptic flow ratio between neutrons and hydrogen isotopes. It is found that $^3$H/$^3$He yield ratio is sensitive to the nuclear symmetry energy at sub-normal densities, while the elliptic flow ratio between neutrons and hydrogen isotopes is sensitive to the high-density behavior of the nuclear symmetry energy. By comparing the UrQMD calculations with 21 Skyrme interactions to the
transverse-momentum dependent elliptic flow ratio $v_{2}^{n}/v_{2}^{H}$, the slope parameter
of the density-dependent symmetry energy is extracted to be $L = 89 \pm 45$~MeV within a 2-$\sigma$
confidence limit. The large uncertainty is partly due to the large error bars in the experimental data and to the fact that the effect of $K_{sym}$ on the elliptic flow ratio has not been disentangled. In the near future, we will concentrate on the investigation of $K_{sym}$ with heavy-ion collisions. The $p_t$-integrated elliptic flow ratio and difference $v_2^{n}-v_2^{p}$, $v_2^{n}-v_2^{H}$, $v_2^{n}/v_2^{p}$, and $v_2^{n}/v_2^{H}$, also can be used to obtain constraints on the slope parameter $L$. Overall, $L$=60-85 MeV is found to be overlapped with the constraints obtained with the UrQMD model and the T\"{u}bingen QMD model by using the data of the elliptic flow ratio (difference) between neutrons and protons (hydrogen isotopes, all charged particles).

Finally, we would like to point out that to achieve a better understanding about the nuclear equation-of-state and the in-medium nucleon-nucleon cross section by using heavy-ion collisions, a detailed and systematical investigation on how the sensitive observables varies with beam energy and collision system is quite necessary. On one hand, the current and future rare isotope beam facilities (e.g., the CSR and the HIAF in China, the FRIB in the United States, the RIBF in Japan, the SPIRAL2 in France, the FAIR in Germany) around the world, will provide more and more experimental data in the next decades, offering new opportunities for
theoretical investigation. On the other hand, uncertainties in transport models, e.g., model-dependent results observed in the comparison of different transport models \cite{Xu:2016lue,Zhang:2017esm,Ono:2019ndq}, need to be understood and solved. Endeavors of both experimentalists and theorists are mandatory to achieve a tight constraint on the nuclear equation-of-state.

To probe the high density behavior of the nuclear symmetry energy with heavy-ion collisions, the $\pi^{-}$/$\pi^{+}$ yield ratio and the elliptic flow ratio between neutrons and protons are two of the most popular observables so far, as the corresponding experimental data are available. As the contribution of the curvature parameter $K_{sym}$ becomes more and more important when studying the high-density behavior of the nuclear symmetry energy, constraint on the $K_{sym}$ is quite necessary\cite{Guo:2018flw,Guo:2019onu}. The effects of the pion and $\Delta$ potentials\cite{Li:2015hfa,Cozma:2016qej,Zhang:2017mps,Liu:2018xvd,Guo:2014fba,Guo:2015tra,Yong:2017cdl}, the in-medium threshold effects on $\Delta$ resonance production and decay\cite{Song:2015hua,Li:2016xix,Zhang:2017mps,Li:2017pis,Cui:2018bkw,Cui:2018qrg,Cui2019}, and other issues\cite{Guo:2014usa,Gao:2018nnp,Cheng:2016pso} on the $\pi^{-}$/$\pi^{+}$ yield ratio need to be understood before a more reliable constraint on the nuclear symmetry energy can be achieved. For the elliptic flow ratio between neutrons and protons, one of a great challenge for experimental techniques is to measure the flow of neutrons with high precision, while from a theoretical point of view, the influence of the neutron-proton effective mass splitting ought to be isolated in advance\cite{Li:2015pma,Li:2018lpy,Xie:2013bsa,Xie:2015xma,Feng:2011pu,Feng:2018emx,feng2012,Gior,Tong}.
In addition, the effects of nucleon-nucleon short-range correlations as well as the associated nucleon momentum distributions in heavy-ion collisions also need to be studied\cite{Li:2018lpy,Hen:2014yfa,Li:2014vua,Liu:2014tqa,Yong:2015gma,Yong:2017zgg,Yang:2018xtl,Yong:2018eeq,Yang:2019jwo}.

Besides using heavy-ion collisions, astrophysical observations such as the mass-radius relation and tidal deformability of neutron stars and gravitational waves also can be used to constrain the density-dependent nuclear symmetry energy \cite{Li:2019xxz,Zhang:2018bwq,Xie:2019sqb,Tsang:2019mlz,Baiotti:2019sew}. Together with constraints on the nuclear equation of state from observables in both nuclear physics (with nuclear structure properties and heavy-ion collisions) and astrophysics (e.g., neutron stars and their mergers), a more precise picture of the nuclear equation of state in a wider density range will be achieved.

\begin{acknowledgments}
We are very grateful to Zhuxia Li, Hongfei Zhang, Chenchen Guo, Arnaud Le F\`evre, Yvonne Leifels, Wolfgang Trautmann for collaborating with us on some of
the topics discussed in this review. The authors acknowledge
support by the computing server C3S2 in Huzhou University. The work is supported in part by the National Natural
Science Foundation of China (Nos. 11875125, 11947410, 11505057), and the Zhejiang Provincial
Natural Science Foundation of China under Grants No. LY18A050002 and No. LY19A050001, and the ``Ten Thousand Talent Program" of Zhejiang province.
\end{acknowledgments}


\begin{thebibliography}{0}

\bibitem{BALi08}
  B.~A.~Li, L.~W.~Chen, C.~M.~Ko,
  Phys.\ Rep.\  {\bf 464}, 113-281 (2008).

  \bibitem{Tsang:2012se}
  M. B. Tsang, J. R. Stone, F. Camera, P. Danielewicz, S. Gandolfi, K. Hebeler, C. J. Horowitz,
  Jenny Lee, W. G. Lynch, Z. Kohley, R. Lemmon, P. M\"{o}ller, T. Murakami, S. Riordan, X. Roca-Maza, F. Sammarruca, A. W. Steiner, I. Vida\~{n}a, and S. J. Yennello,
  Phys.\ Rev.\ C {\bf 86}, 015803 (2012).


\bibitem{Baldo:2016jhp}
  M.~Baldo and G.~F.~Burgio,
  Prog.\ Part.\ Nucl.\ Phys.\  {\bf 91}, 203 (2016).

\bibitem{Oertel:2016bki}
  M.~Oertel, M.~Hempel, T.~Kl\"{a}hn and S.~Typel,
  Rev.\ Mod.\ Phys.\  {\bf 89}, no. 1, 015007 (2017).

\bibitem{Li:2018lpy}
  B.~A.~Li, B.~J.~Cai, L.~W.~Chen and J.~Xu,
  Prog.\ Part.\ Nucl.\ Phys.\  {\bf 99}, 29 (2018).

\bibitem{Roca-Maza:2018ujj}
  X.~Roca-Maza and N.~Paar,
  Prog.\ Part.\ Nucl.\ Phys.\  {\bf 101}, 96 (2018).

\bibitem{Burrello:2019wyi}
  S.~Burrello, M.~Colonna and H.~Zheng,
  Front.\ in Phys.\  {\bf 7}, 53 (2019).

\bibitem{Giuliani:2013ppnp}
  G.~Giuliani, H.~Zheng and A.~Bonasera,
  Prog.\ Part.\ Nucl.\ Phys.\  {\bf 76}, 116 (2014).
\bibitem{Ma:2018wtw}
  C.~W.~Ma and Y.~G.~Ma,
  Prog.\ Part.\ Nucl.\ Phys.\  {\bf 99}, 120 (2018).
\bibitem{Ono:2019jxm}
  A.~Ono,
  Prog.\ Part.\ Nucl.\ Phys.\  {\bf 105}, 139 (2019).

\bibitem{Xu:2019hqg}
  J.~Xu,
  Prog.\ Part.\ Nucl.\ Phys.\  {\bf 106}, 312 (2019).
\bibitem{Gao:2019vby}
  H.~Gao, S.~K.~Ai, Z.~J.~Cao, B.~Zhang, Z.~Y.~Zhu, A.~Li, N.~B.~Zhang and A.~Bauswein,
  Front.\ Phys.\ (Beijing) {\bf 15}, no. 2, 24603 (2020).


\bibitem{Chen:2013uua}
  L.~W.~Chen, C.~M.~Ko, B.~A.~Li, C.~Xu and J.~Xu,
  Eur.\ Phys.\ J.\ A {\bf 50}, 29 (2014).
\bibitem{Xu:2009vi}
  J.~Xu, L.~W.~Chen, B.~A.~Li and H.~R.~Ma,
  Astrophys.\ J.\  {\bf 697}, 1549 (2009).

\bibitem{Cai:2011zn}
  B.~J.~Cai and L.~W.~Chen,
  Phys.\ Rev.\ C {\bf 85}, 024302 (2012).

\bibitem{Steiner:2006bx}
  A.~W.~Steiner,
  Phys.\ Rev.\ C {\bf 74}, 045808 (2006).

\bibitem{Pu:2017kjx}
  J.~Pu, Z.~Zhang and L.~W.~Chen,
  Phys.\ Rev.\ C {\bf 96}, no. 5, 054311 (2017).

\bibitem{Liu:2018far}
  Z.~W.~Liu, Z.~Qian, R.~Y.~Xing, J.~R.~Niu and B.~Y.~Sun,
  Phys.\ Rev.\ C {\bf 97}, no. 2, 025801 (2018).




  \bibitem{Bass98}S. A. Bass {\it et al.}, [UrQMD-Collaboration], Prog. Part. Nucl. Phys. {\bf 41}, 255 (1998).

\bibitem{Bleicher:1999xi}
  M.~Bleicher, E.~Zabrodin, C.~Spieles, S.~A.~Bass, C.~Ernst, S.~Soff, L.~Bravina, M.~Belkacem {\it et al.},
  J.\ Phys.\ G {\bf 25}, 1859 (1999).

\bibitem{Li:2011zzp}
  Q.~Li, C.~Shen, C.~Guo, Y.~Wang, Z.~Li, J.~{\L}ukasik, W.~Trautmann,
  Phys.\ Rev.\  {\bf C83}, 044617 (2011).

\bibitem{Li:2012ta}
  Q.~Li, G.~Graf and M.~Bleicher,
  Phys.\ Rev.\ C {\bf 85}, 034908 (2012).

\bibitem{Aichelin:1991xy} J. Aichelin, Phys. Rept.  {\bf 202}, 233 (1991).

  \bibitem{Hartnack:1997ez} C. Hartnack, R. K. Puri, J. Aichelin, J. Konopka, S. A. Bass, H. Stoecker and W. Greiner,  Eur. Phys. J.  A {\bf 1}, 151 (1998).
\bibitem{Li:2005gfa} Q. F. Li, Z. X. Li, S. Soff, M. Bleicher and H. Stoecker,  J.\ Phys.\ G {\bf 32}, 151 (2006).

\bibitem{Zhang:2018rle}
  F.~S.~Zhang, C.~Li, L.~Zhu and P.~Wen,
  Front.\ Phys.\ (Beijing) {\bf 13}, no. 6, 132113 (2018).

\bibitem{Zhang:2006vb}
  Y.~Zhang and Z.~Li,
  Phys.\ Rev.\ C {\bf 74}, 014602 (2006).


\bibitem{Zhang:2007gd}
  Y.~Zhang, Z.~Li and P.~Danielewicz,
Phys.\ Rev.\ C {\bf 75}, 034615 (2007); Y. X. Zhang, Z. X. Li and P. Danielewicz, Phys. Rev. C {\bf 75}, 034615 (2007).

  \bibitem{Wang:2013wca}
  Y.~Wang, C.~Guo, Q.~Li, H.~Zhang, Z.~Li, and W.~Trautmann,
  Phys.\ Rev.\ C {\bf 89}, 034606 (2014).

\bibitem{Wang:2014rva}
  Y.~Wang, C.~Guo, Q.~Li, H.~Zhang, Y.~Leifels and W.~Trautmann,
  Phys.\ Rev.\ C {\bf 89}, no. 4, 044603 (2014).


\bibitem{Du:2018ruo}
  Y.~Du, Y.~Wang, Q.~Li and L.~Liu,
  Sci.\ China Phys.\ Mech.\ Astron.\  {\bf 61}, no. 6, 062011 (2018).
\bibitem{Liu:2018xvd}
  Y.~Liu, Y.~Wang, Q.~Li and L.~Liu,
  Phys.\ Rev.\ C {\bf 97}, no. 3, 034602 (2018).

   \bibitem{lgq14} G. Q. Li and R. Machleidt, Phys. Rev. C {\bf 48}, 1702 (1993); G. Q. Li and R. Machleidt, Phys. Rev. C {\bf49}, 566 (1994).

\bibitem{Sammarruca:2005tk} F. Sammarruca and P. Krastev, Phys. Rev. C {\bf 73}, 014001 (2006).

\bibitem{HJS} H. J. Schulze, A. Schnell, G. Ropke, and U. Lombardo, Phys. Rev. C {\bf55}, 3006 (1997).

\bibitem{CF} C. Fuchs, A. Faessler, and M. El-Shabshiry, Phys. Rev. C {\bf64}, 024003 (2001).

\bibitem{HFZ07} H. F. Zhang, Z. H. Li, U. Lombardo, P. Y. Luo, F. Sammarruca, and W. Zuo, Phys. Rev. C {\bf76}, 054001 (2007); H. F. Zhang, U. Lombardo, and W. Zuo, Phys. Rev. C {\bf82}, 015805 (2010).

\bibitem{WGL} W. G. Love and M. A. Franey, Phys. Rev. C {\bf24}, 3 (1981).

\bibitem{Alm:1995chb} T. Alm, G. R\"{o}pke, W. Bauer, F. Daffin and M. Schmidt, Nucl. Phys. A {\bf 587}, 815 (1995).


\bibitem{Mao:1994zza} G. J. Mao, Z. X. Li, Y. Z. Zhuo and Z. Q. Yu, Phys. Lett. B {\bf 327}, 183 (1994); G. J. Mao, Z. X. Li, Y. Z. Zhuo, Y. Han and Z. Yu, Phys. Rev. C {\bf 49}, 3137 (1994).

\bibitem{Li:2003vd} Q. F. Li, Z. X. Li and G. J. Mao, Phys. Rev. C {\bf 62}, 014606 (2000); Q. F. Li, Z. X. Li and E. G. Zhao, Phys. Rev. C {\bf 69}, 017601 (2004); Q. F. Li and Z. X. Li, Phys. Lett. B {\bf 773}, 557 (2017).



\bibitem{gd10} G. D. Westfall $et al$., Phys. Rev. Lett. {\bf71}, 1986 (1993).
\bibitem{dds} D. D. S. Coupland, W. G. Lynch, M. B. Tsang, P. Danielewicz, Y. X. Zhang, Phys. Rev. C {\bf84}, 054603 (2011).
\bibitem{Li:2005iba} B. A. Li, P. Danielewicz and W. G. Lynch, Phys. Rev. C {\bf 71}, 054603 (2005); B. A. Li and L. W. Chen, Phys. Rev. C {\bf 72}, 064611 (2005).
\bibitem{Feng:2011eu} Z. Q. Feng, Phys. Rev. C {\bf 85}, 014604 (2012).
\bibitem{Guo:2013fka}
  W.~M.~Guo, G.~C.~Yong, Y.~Wang, Q.~Li, H.~Zhang and W.~Zuo,
  Phys.\ Lett.\ B {\bf 726}, 211 (2013).

\bibitem{Bass:1995pj}
  S.~A.~Bass, C.~Hartnack, H.~Stoecker and W.~Greiner,
  Phys.\ Rev.\ C {\bf 51} (1995) 3343.

\bibitem{Wang:2012sy}
  Y.~Wang, C.~Guo, Q.~Li and H.~Zhang,
  Sci.\ China Phys.\ Mech.\ Astron.\  {\bf 55}, 2407 (2012).

\bibitem{Guo:2012aa}
 C.~Guo,  Y.~Wang, Q.~Li, W.~Trautmann, L.~Liu and L.~Wu,
  Sci.\ China Phys.\ Mech.\ Astron.\  {\bf 55}, 252 (2012).
  \bibitem{lipc}
 P. C. Li, Y. J.  Wang, Q. F. Li, H. F. Zhang,
 Phys. Rev. C {\bf97}, 044620 (2018). 
\bibitem{Li:2018bus}
  P.~C.~Li, Y.~J.~Wang, Q.~F.~Li and H.~F.~Zhang,
  Nucl.\ Sci.\ Tech.\  {\bf 29}, no. 12, 177 (2018).
\bibitem{Russotto:2011hq}
  P.~Russotto {\it et al.},
  Phys.\ Lett.\ B {\bf 697}, 471 (2011).
\bibitem{Zhang:2012qm}
  Y.~Zhang, Z.~Li, C.~Zhou, M.~B.~Tsang,
  Phys.\ Rev.\ C {\bf 85}, 051602 (2012).
\bibitem{Zbiri:2006ts}
  K.~Zbiri {\it et al.},
  Phys.\ Rev.\ C {\bf 75}, 034612 (2007).
\bibitem{Li:2016mqd}
  Q.~Li, Y.~Wang, X.~Wang and C.~Shen,
  Sci.\ China Phys.\ Mech.\ Astron.\  {\bf 59}, no. 2, 622001 (2016).

\bibitem{Reisdorf:1997fx}
  W.~Reisdorf and H.~G.~Ritter,
  Ann.\ Rev.\ Nucl.\ Part.\ Sci.\  {\bf 47}, 663 (1997).


\bibitem{FOPI:2011aa}
  W.~Reisdorf {\it et al.}  [FOPI Collaboration],
  Nucl.\ Phys.\ A {\bf 876}, 1 (2012).
\bibitem{Heinz:2013th}
  U.~Heinz and R.~Snellings,
  Ann.\ Rev.\ Nucl.\ Part.\ Sci.\  {\bf 63}, 123 (2013).



  \bibitem{Danie02} P. Danielewicz, R. Lacey, and W. G.Lynch, Science 298,1592 (2002).
\bibitem{Ollitrault:1997vz}
  J.~Y.~Ollitrault,
  Nucl.\ Phys.\ A {\bf 638}, 195 (1998).

\bibitem{Andronic:2006ra}
  A.~Andronic, J.~Lukasik, W.~Reisdorf and W.~Trautmann,
  Eur.\ Phys.\ J.\ A {\bf 30}, 31 (2006).
\bibitem{Andronic:2004cp}
  A.~Andronic {\it et al.} [FOPI Collaboration],
  Phys.\ Lett.\ B {\bf 612}, 173 (2005).
  \bibitem{Le}A. Le F\`{e}vre, Y. Leifels, C. Hartnack, and J. Aichelin, Phys. Rev. C {\bf 98}, 034901 (2018).


\bibitem{Zheng:1999gt}
  Y.~M.~Zheng, C.~M.~Ko, B.~A.~Li and B.~Zhang,
  Phys.\ Rev.\ Lett.\  {\bf 83}, 2534 (1999).
\bibitem{Persram:2001dg}
  D.~Persram and C.~Gale,
  Phys.\ Rev.\ C {\bf 65} (2002) 064611.
\bibitem{Gaitanos:2004ic}
  T.~Gaitanos, C.~Fuchs and H.~H.~Wolter,
  Phys.\ Lett.\ B {\bf 609}, 241 (2005).
\bibitem{Li:2005jy}
  B.~A.~Li and L.~W.~Chen,
  Phys.\ Rev.\ C {\bf 72}, 064611 (2005).
\bibitem{Kaur:2016eaf}
  M.~Kaur and S.~Gautam,
  J.\ Phys.\ G {\bf 43}, no. 2, 025103 (2016).
\bibitem{Barker:2016hqv}
  B.~Barker and P.~Danielewicz,
  Phys.\ Rev.\ C {\bf 99} (2019) no.3,  034607.
\bibitem{Basrak:2016cbo}
  Z.~Basrak, P.~Eudes and V.~de la Mota,
  Phys.\ Rev.\ C {\bf 93} (2016) no.5,  054609.

\bibitem{Lehaut:2010zz}
  G.~Lehaut {\it et al.} [INDRA and ALADIN Collaborations],
  Phys.\ Rev.\ Lett.\  {\bf 104} (2010) 232701.
\bibitem{Reisdorf:2004wg}
  W.~Reisdorf {\it et al.} [FOPI Collaboration],
  Phys.\ Rev.\ Lett.\  {\bf 92}, 232301 (2004).

\bibitem{FOPI:2010aa}
  W.~Reisdorf {\it et al.}  [FOPI Collaboration],
  Nucl.\ Phys.\ A {\bf 848}, 366 (2010).

\bibitem{lipcjpg}
  P.~Li, Y.~Wang, Q.~Li, J.~Wang and H.~Zhang,
 J.\ Phys.\ G {\bf 47}, 035108 (2020).

\bibitem{Wang:2016yti}
  Y.~Wang, C.~Guo, Q.~Li, Z.~Li, J.~Su and H.~Zhang,
  Phys.\ Rev.\ C {\bf 94}, no. 2, 024608 (2016).



\bibitem{Stone:2014wza}
  J.~R.~Stone, N.~J.~Stone and S.~A.~Moszkowski,
  Phys.\ Rev.\ C {\bf 89}, no. 4, 044316 (2014).


\bibitem{Khan:2013mga}
  E.~Khan and J.~Margueron,
  Phys.\ Rev.\ C {\bf 88}, no. 3, 034319 (2013); E.~Khan and J.~Margueron,
  Phys.\ Rev.\ Lett.\  {\bf 109}, 092501 (2012).

\bibitem{Molitoris:1986pp}
  J.~J.~Molitoris, D.~Hahn and H.~Stoecker,
  Nucl.\ Phys.\ A {\bf 447}, 13c (1986).

\bibitem{Molitoris:1985gs}
  J.~J.~Molitoris and H.~St\"{o}cker,
  Phys.\ Rev.\ C {\bf 32}, 346 (1985).
\bibitem{Kruse:1985hy}
  H.~Kruse, B.~V.~Jacak and H.~St\"{o}cker,
  Phys.\ Rev.\ Lett.\  {\bf 54}, 289 (1985).

\bibitem{Aichelin:1986ss}
  J.~Aichelin and C.~M.~Ko,
  Phys.\ Rev.\ Lett.\  {\bf 55}, 2661 (1985).
\bibitem{Stoecker:1986ci}
  H.~St\"{o}cker and W.~Greiner,
  Phys.\ Rept.\  {\bf 137}, 277 (1986).
\bibitem{Cassing:1990dr}
  W.~Cassing, V.~Metag, U.~Mosel and K.~Niita,
  Phys.\ Rept.\  {\bf 188}, 363 (1990).

\bibitem{Hartnack:2005tr}
 C.~Sturm {\it et al.},
  Phys.\ Rev.\ Lett.\  {\bf 86}, 39 (2001);
  C.~Fuchs, A.~Faessler, E.~Zabrodin and Y.~-M.~Zheng,
  Phys.\ Rev.\ Lett.\  {\bf 86}, 1974 (2001);
  C. Hartnack, H.~Oeschler and J.~Aichelin,
  Phys.\ Rev.\ Lett.\  {\bf 96}, 012302 (2006).

\bibitem{Feng:2011dp}
  Z.~Q.~Feng,
  Phys.\ Rev.\ C {\bf 83}, 067604 (2011).


\bibitem{Fevre:2015fza}
  A.~Le F\`evre, Y.~Leifels, W.~Reisdorf, J.~Aichelin and C.~Hartnack,
  Nucl.\ Phys.\ A {\bf 945}, 112 (2016).

\bibitem{Xu:2016lue}
  Jun Xu, Lie-Wen Chen, ManYee Betty Tsang {\it et al.},
  Phys.\ Rev.\ C {\bf 93}, no. 4, 044609 (2016).


     \bibitem{Dutra:2012mb}
  M.~Dutra, O.~Lourenco, J.~S.~Sa Martins, A.~Delfino, J.~R.~Stone and P.~D.~Stevenson,
  Phys.\ Rev.\ C {\bf 85}, 035201 (2012).



\bibitem{Wang:2018hsw}
  Y.~Wang, C.~Guo, Q.~Li, A.~Le F¨¨vre, Y.~Leifels and W.~Trautmann,
  Phys.\ Lett.\ B {\bf 778}, 207 (2018).
\bibitem{Russotto:2016ucm}
  P.~Russotto {\it et al.},
  Phys.\ Rev.\ C {\bf 94}, no. 3, 034608 (2016).


\bibitem{Chen:2011ib}
  L.~W.~Chen,
  Sci.\ China Phys.\ Mech.\ Astron.\  {\bf 54}, 124 (2011).
\bibitem{Chen:2009wv}
  L.~W.~Chen, B.~J.~Cai, C.~M.~Ko, B.~A.~Li, C.~Shen and J.~Xu,
  Phys.\ Rev.\ C {\bf 80}, 014322 (2009).




\bibitem{Li:2002qx}
  B.~A.~Li,
  Phys.\ Rev.\ Lett.\  {\bf 88}, 192701 (2002).


\bibitem{Xiao:2009zza}
  Z.~Xiao, B.~A.~Li, L.~W.~Chen, G.~C.~Yong and M.~Zhang,
  Phys.\ Rev.\ Lett.\  {\bf 102}, 062502 (2009).

\bibitem{Feng:2009am}
  Z.~Q.~Feng and G.~M.~Jin,
  Phys.\ Lett.\ B {\bf 683}, 140 (2010).

\bibitem{Zhang:2018bwq}
  N.~B.~Zhang and B.~A.~Li,
  Eur.\ Phys.\ J.\ A {\bf 55}, 39 (2019).
\bibitem{Xie:2019sqb}
  W.~J.~Xie and B.~A.~Li,
  Astrophys.\ J.\  {\bf 883}, 174 (2019).
\bibitem{zhouchen}
Y. Zhou, L. Chen, and Z. Zhang, Phys. Rev. D {\bf 99}, 121301 (2019).



\bibitem{Wang:2014aba}
  Y.~Wang, C.~Guo, Q.~Li and H.~Zhang,
  Eur.\ Phys.\ J.\ A {\bf 51}, no. 3, 37 (2015).



\bibitem{DiToro:2010ku}
  M.~Di Toro, V.~Baran, M.~Colonna and V.~Greco,
  J.\ Phys.\ G {\bf 37}, 083101 (2010).

\bibitem{Tsang:2008fd}
  M.~B.~Tsang, Y.~Zhang, P.~Danielewicz, M.~Famiano, Z.~Li, W.~G.~Lynch and A.~W.~Steiner,
  Phys.\ Rev.\ Lett.\  {\bf 102}, 122701 (2009);
  M.~B.~Tsang {\it et al.}, Int.\ J.\ Mod.\ Phys.\ E {\bf 19}, 1631 (2010).

\bibitem{Lopez:2007}
  X.~Lopez, Y.~J. Kim, N.~Herrmann, A.~Andronic, V.~Barret, Z.~Basrak, N.~Bastid, M.~L.~Benabderrahmane
 {\it et al.},
  Phys.\ Rev.\ C\  {\bf 75}, 011901(R) (2007).



\bibitem{Cozma:2011nr}
  M.~D.~Cozma,
  Phys.\ Lett.\ B {\bf 700}, 139 (2011).

\bibitem{Cozma:2013sja}
  M.~D.~Cozma, Y.~Leifels, W.~Trautmann, Q.~Li and P.~Russotto,
  Phys.\ Rev.\ C {\bf 88}, 044912 (2013).

\bibitem{Kumar:2011td}
  S.~Kumar, Y.~G.~Ma, G.~Q.~Zhang and C.~L.~Zhou,
  Phys.\ Rev.\ C {\bf 85}, 024620 (2012).
\bibitem{Lu:2016htm}
  L.~L\"{u}, H.~Yi, Z.~Xiao, M.~Shao, S.~Zhang, G.~Xiao and N.~Xu,
  Sci.\ China Phys.\ Mech.\ Astron.\  {\bf 60}, no. 1, 012021 (2017).

\bibitem{Russotto:2013fza}
  P.~Russotto, M.~D.~Cozma, A.~Fevre, Y.~Leifels, R.~Lemmon, Q.~Li, J.~Lukasik and W.~Trautmann,
  Eur.\ Phys.\ J.\ A {\bf 50}, 38 (2014).

\bibitem{Xie:2013np}
  W.~J.~Xie, J.~Su, L.~Zhu and F.~S.~Zhang,
  Phys.\ Lett.\ B {\bf 718}, 1510 (2013).

\bibitem{Li:2005zza}
  Q.~Li, Z.~Li, S.~Soff, R.~K.~Gupta, M.~Bleicher and H.~St\"{o}cker,
  J.\ Phys.\ G {\bf 31}, 1359 (2005).

\bibitem{LI:2005zi}
  Q.~Li, Z.~Li, E.~Zhao and R.~K.~Gupta,
  Phys.\ Rev.\ C {\bf 71}, 054907 (2005).

\bibitem{Gautam:2010da}
  S.~Gautam, A.~D.~Sood, R.~K.~Puri and J.~Aichelin
  Phys.\ Rev.\ C {\bf 83}, 014603 (2011).
\bibitem{Tsang:2016foy}
  M.~B.~Tsang {\it et al.} [S¦ÐRIT Collaboration],
  Phys.\ Rev.\ C {\bf 95}, no. 4, 044614 (2017).



   \bibitem{Li:2012mw}
  B.~A.~Li, L.~W.~Chen, F.~J.~Fattoyev, W.~G.~Newton and C.~Xu,
  J.\ Phys.\ Conf.\ Ser.\  {\bf 413}, 012021 (2013).

\bibitem{Chen:2012pk}
  L.~W.~Chen,
  arXiv:1212.0284 [nucl-th].

  \bibitem{Wolter}
H. Wolter,
Proceedings of Science (Bormio2012), 059 (2012).

\bibitem{Li:2013ola}
  B.~A.~Li and X.~Han,
  Phys.\ Lett.\ B {\bf 727}, 276 (2013).

\bibitem{Danielewicz:2013upa}
  P.~Danielewicz, J.~Lee,
  Nucl.\ Phys.\ A {\bf 922}, 1 (2014).

\bibitem{RocaMaza:2012mh}
  X.~Roca-Maza {\it et al.},
  Phys.\ Rev.\ C {\bf 87}, 034301 (2013).

\bibitem{brown}
  B.~A.~Brown, Phys.\ Rev.\ Lett. {\bf 111}, 232502 (2013).

\bibitem{Zhang:2013wna}
  Z.~Zhang, L.~-W.~Chen,
  Phys.\ Lett.\ B {\bf 726}, 234 (2013).

\bibitem{Wang:2015kof}
  N.~Wang, M.~Liu, L.~Ou, and Y.~Zhang,
  Phys.\ Lett.\ B {\bf 751}, 553 (2015).

\bibitem{Fan:2014rha}
  X.~Fan, J.~Dong, W.~Zuo,
  Phys.\ Rev.\ C {\bf 89}, 017305 (2014).


   \bibitem{Chen:2003qj}
  L.~-W.~Chen, C.~M.~Ko, B.~-A.~Li,
  Phys.\ Rev.\ C {\bf 68}, 017601 (2003); L.~-W.~Chen, C.~M.~Ko, B.~-A.~Li,
  Nucl.\ Phys.\ A {\bf 729}, 809 (2003).
\bibitem{Chen:2004kj}
  L.~-W.~Chen, C.~M.~Ko, B.~-A.~Li,
  Phys.\ Rev.\ C {\bf 69}, 054606 (2004).


\bibitem{Li:2005kqa}
  Q.~Li, Z.~Li, S.~Soff, M.~Bleicher, H.~St\"{o}cker,
  Phys.\ Rev.\ C {\bf 72}, 034613 (2005).
\bibitem{Zhang:2005sm}
  Y.~Zhang, Z.~Li,
  Phys.\ Rev.\ C {\bf 71}, 024604 (2005).
\bibitem{Yong:2009te}
  G.~-C.~Yong, B.~-A.~Li, L.~-W.~Chen, X.~-C.~Zhang,
  Phys.\ Rev.\ C {\bf 80}, 044608 (2009).


\bibitem{Li:2000bj}
  B.~A.~Li,
  Phys.\ Rev.\ Lett.\  {\bf 85}, 4221 (2000).


\bibitem{Greco:2002sp}
  V.~Greco, V.~Baran, M.~Colonna, M.~Di Toro, T.~Gaitanos and H.~H.~Wolter,
  Phys.\ Lett.\ B {\bf 562}, 215 (2003).


\bibitem{Trautmann:2009kq}
  W.~Trautmann {\it et al.},
  Prog.\ Part.\ Nucl.\ Phys.\  {\bf 62}, 425 (2009).

\bibitem{Trautmann:2010at}
  W.~Trautmann {\it et al.},
  Int.\ J.\ Mod.\ Phys.\ E {\bf 19}, no. 08n09, 1653 (2010).

\bibitem{Cozma:2017bre}
  M.~D.~Cozma,
  Eur.\ Phys.\ J.\ A {\bf 54}, no. 3, 40 (2018).

\bibitem{wyj-plb2}
  Y.~Wang, Q.~Li, A.~Le F¨¨vre, and Y.~Leifels,
  Phys.\ Lett.\ B {\bf 802}, 135249 (2020).

\bibitem{Zhang:2017esm}
  Y.~X.~Zhang {\it et al.},
  Phys.\ Rev.\ C {\bf 97}, no. 3, 034625 (2018).

\bibitem{Ono:2019ndq}
  A.~Ono {\it et al.},
  Phys.\ Rev.\ C {\bf 100}, no. 4, 044617 (2019).


\bibitem{Guo:2018flw}
  Y.~F.~Guo and G.~C.~Yong,
  Phys.\ Rev.\ C {\bf 100}, no. 1, 014617 (2019).

\bibitem{Guo:2019onu}
  Y.~F.~Guo and G.~C.~Yong,
  arXiv:1909.13566 [nucl-th].




\bibitem{Li:2015hfa}
  B.~A.~Li,
  Phys.\ Rev.\ C {\bf 92}, no. 3, 034603 (2015).

\bibitem{Cozma:2016qej}
  M.~D.~Cozma,
  Phys.\ Rev.\ C {\bf 95}, no. 1, 014601 (2017).


\bibitem{Zhang:2017mps}
  Z.~Zhang and C.~M.~Ko,
  Phys.\ Rev.\ C {\bf 95}, no. 6, 064604 (2017).


\bibitem{Guo:2014fba}
  W.~M.~Guo, G.~C.~Yong, H.~Liu and W.~Zuo,
  Phys.\ Rev.\ C {\bf 91}, no. 5, 054616 (2015).

\bibitem{Guo:2015tra}
  W.~M.~Guo, G.~C.~Yong and W.~Zuo,
  Phys.\ Rev.\ C {\bf 92}, no. 5, 054619 (2015).
\bibitem{Yong:2017cdl}
  G.~C.~Yong,
  Phys.\ Rev.\ C {\bf 96}, no. 4, 044605 (2017).

\bibitem{Song:2015hua}
  T.~Song and C.~M.~Ko,
  Phys.\ Rev.\ C {\bf 91}, no. 1, 014901 (2015).
\bibitem{Li:2016xix}
  Q.~Li and Z.~Li,
  Phys.\ Lett.\ B {\bf 773}, 557 (2017).
\bibitem{Li:2017pis}
  Q.~Li and Z.~Li,
  Sci.\ China Phys.\ Mech.\ Astron.\  {\bf 62}, no. 7, 972011 (2019).
\bibitem{Cui:2018bkw}
  Y.~Cui, Y.~Zhang and Z.~Li,
  Chin.\ Phys.\ C {\bf 43}, no. 2, 024105 (2019).
  \bibitem{Cui2019}
  Y.~Cui, Y.~Zhang and Z.~Li,
  Chin.\ Phys.\ C {\bf 44}, no. 2, 024106 (2020).
\bibitem{Cui:2018qrg}
  Y.~Cui, Y.~Zhang and Z.~Li,
  Phys.\ Rev.\ C {\bf 98}, no. 5, 054605 (2018).
\bibitem{Guo:2014usa}
  W.~M.~Guo, G.~C.~Yong and W.~Zuo,
  Phys.\ Rev.\ C {\bf 90}, no. 4, 044605 (2014).
\bibitem{Gao:2018nnp}
  Y.~Gao, G.~C.~Yong, L.~Zhang and W.~Zuo,
  Phys.\ Rev.\ C {\bf 97}, no. 1, 014609 (2018).
\bibitem{Cheng:2016pso}
  S.~J.~Cheng, G.~C.~Yong and D.~H.~Wen,
  Phys.\ Rev.\ C {\bf 94}, no. 6, 064621 (2016).


\bibitem{Li:2015pma}
  B.~A.~Li and L.~W.~Chen,
  Mod.\ Phys.\ Lett.\ A {\bf 30}, no. 13, 1530010 (2015).

\bibitem{Xie:2013bsa}
  W.~J.~Xie, J.~Su, L.~Zhu and F.~S.~Zhang,
  Phys.\ Rev.\ C {\bf 88}, no. 6, 061601 (2013).
\bibitem{Xie:2015xma}
  W.~J.~Xie, Z.~Q.~Feng, J.~Su and F.~S.~Zhang,
  Phys.\ Rev.\ C {\bf 91}, no. 5, 054609 (2015).


\bibitem{Feng:2011pu}
  Z.~Q.~Feng,
  Phys.\ Rev.\ C {\bf 84}, 024610 (2011).
\bibitem{Feng:2018emx}
  Z.~Q.~Feng,
  Nucl.\ Sci.\ Tech.\  {\bf 29}, no. 3, 40 (2018).
  \bibitem{feng2012}Z. Q. Feng, Nucl. Phys. A {\bf 878}, 3 (2012).

  \bibitem{Gior} V. Giordano, M. Colonna, M. Di Toro, V. Greco, J. Rizzo, Phys. Rev. C {\bf 81} 044611 (2010).
\bibitem{Tong} L. Y. Tong, P. C. Li, F. P. Li, Y. J. Wang, Q. F. Li, F. X. Liu, Chin.\ Phys.\ C {\bf **}, no. *, ****** (2020). Accepted.
\bibitem{Hen:2014yfa}
  O.~Hen, B.~A.~Li, W.~J.~Guo, L.~B.~Weinstein and E.~Piasetzky,
  Phys.\ Rev.\ C {\bf 91}, no. 2, 025803 (2015).

\bibitem{Li:2014vua}
  B.~A.~Li, W.~J.~Guo and Z.~Shi,
  Phys.\ Rev.\ C {\bf 91}, no. 4, 044601 (2015).

\bibitem{Liu:2014tqa}
  H.~L.~Liu, G.~C.~Yong and D.~H.~Wen,
  Phys.\ Rev.\ C {\bf 91}, no. 2, 024604 (2015).

\bibitem{Yong:2015gma}
  G.~C.~Yong,
  Phys.\ Lett.\ B {\bf 765}, 104 (2017).


\bibitem{Yong:2017zgg}
  G.~C.~Yong and B.~A.~Li,
  Phys.\ Rev.\ C {\bf 96}, no. 6, 064614 (2017).


\bibitem{Yang:2018xtl}
  Z.~X.~Yang, X.~H.~Fan, G.~C.~Yong and W.~Zuo,
  Phys.\ Rev.\ C {\bf 98}, no. 1, 014623 (2018).

\bibitem{Yong:2018eeq}
  G.~C.~Yong,
  Phys.\ Lett.\ B {\bf 776}, 447 (2018).

\bibitem{Yang:2019jwo}
  Z.~X.~Yang, X.~L.~Shang, G.~C.~Yong, W.~Zuo and Y.~Gao,
  Phys.\ Rev.\ C {\bf 100}, no. 5, 054325 (2019).

\bibitem{Li:2019xxz}
  B.~A.~Li, P.~G.~Krastev, D.~H.~Wen and N.~B.~Zhang,
  Eur.\ Phys.\ J.\ A {\bf 55}, no. 7, 117 (2019).

\bibitem{Tsang:2019mlz}
  M.~B.~Tsang, W.~G.~Lynch, P.~Danielewicz and C.~Y.~Tsang,
  Phys.\ Lett.\ B {\bf 795}, 533 (2019).
\bibitem{Baiotti:2019sew}
  L.~Baiotti,
  Prog.\ Part.\ Nucl.\ Phys.\  {\bf 109}, 103714 (2019).


\end{thebibliography}
\end{document}